\documentclass[aps,prd,twocolumn,twoside, superscriptaddress,tightenlines,nofootinbib,floats, amsmath]{revtex4-1}
\pdfoutput=1
\usepackage{graphicx}
\usepackage{dcolumn}
\usepackage{bm}
\usepackage{appendix}
\usepackage{amsfonts}
\usepackage{comment}
\usepackage[bookmarks=false,colorlinks]{hyperref}
\hypersetup{urlcolor=blue, citecolor=blue}
\usepackage[utf8]{inputenc}
\usepackage{scalerel}
\usepackage{slashed}
\usepackage{color}
\usepackage{mathrsfs}
\usepackage{epsfig}
\usepackage{multirow}
\usepackage{lipsum}
\usepackage{mathrsfs}
\usepackage{enumerate}

\def\beq{\begin{equation}}
\def\eeq{\end{equation}}
\def\bea{\begin{eqnarray}}
\def\eea{\end{eqnarray}}
\def\nn{\nonumber}
\def\nl{\nn \\}
\def\cB{\mathcal{B}}

\usepackage{tabularx}
\usepackage{url}
\usepackage{footmisc}
\usepackage{amsfonts}
\usepackage{cancel}
\usepackage{color}
\usepackage{pifont}
\usepackage{epsfig}
\usepackage{epstopdf}
\usepackage{comment}
\usepackage{booktabs} 
\usepackage{array}
\usepackage{mathrsfs}
\usepackage{ulem}
\usepackage{bbold}
\usepackage{enumitem}
\usepackage{multirow}
\usepackage{cleveref}
\usepackage{upgreek}
\usepackage{xspace}				

\usepackage{braket}
\usepackage{bm}
\def\nn{\nonumber}
\def\nl{\nonumber\\}

\usepackage{epsfig}

\newcommand{\bctaunutau}{b \to c \tau^- {\bar\nu}_\tau}

\def\BDtaunu{\bar{B} \to D^+ \tau^{-} \bar{\nu_\tau}}

\def\RD{R({D^{(*)}})}

\def \s{\sqrt{2}}

\def \cB{{\cal B}}
\def \cL{{\cal L}}

\def \be{\beta}

\def\beq{\begin{equation}}
\def\eeq{\end{equation}}
\def\bea{\begin{eqnarray}}
\def\eea{\end{eqnarray}}
\def\nn{\nonumber}
\def\bit{\begin{itemize}}
\def\eit{\end{itemize}}
\def\roughly#1{\mathrel{\raise.3ex\hbox
{$#1$\kern-.75em\lower1ex\hbox{$\sim$}}}}

\def\sss{\scriptscriptstyle}
\def\.{\!\cdot\!}    \def\:{\cdots}   \def\[{\left[}   \def\]{\right]}
\def\({\left(} \def\){\right)} 

\def\barpk{{\raise.35ex\hbox
{${\sss (}$}}--{\raise.35ex\hbox{${\sss )}$}}}
\def\bbarp{\hbox{$B$\kern-0.9em\raise1.4ex\hbox{\barpk}}}
\def\beq{\begin{equation}}
\def\eeq{\end{equation}}
\def\bea{\begin{eqnarray}}
\def\eea{\end{eqnarray}}
\def\ber{\begin{eqnarray*}}
\def\eer{\end{eqnarray*}}
\def\nn{\nonumber}
\def\sss{\scriptscriptstyle}
\def\roughly#1{\mathrel{\raise.3ex\hbox
{$#1$\kern-.75em\lower1ex\hbox{$\sim$}}}}

\def \lbt{\Lambda_b \to \Lambda_c^+ \tau^- \bar{\nu}_{\tau}}
\def \lbl{\Lambda_b \to \Lambda_c^+ \ell^- \bar{\nu}_{\ell}}

\def\bwt{\begin{widetext}}
\def\ewt{\end{widetext}}
\def\nn{\nonumber}
\def\roughly#1{\mathrel{\raise.3ex\hbox
{$#1$\kern-.75em\lower1ex\hbox{$\sim$}}}}
\def\order{\lower 1.8ex \hbox{\LARGE\~{}}}

\def\BDtaunu{\bar{B} \to D^+ \tau^{-} {\bar\nu}_\tau}

\def\be{\begin{equation}}
\def\ee{\end{equation}}
\def\bea{\begin{eqnarray}}
\def\eea{\end{eqnarray}}

\def\A0{{\cal{A}}_{0}}

\def\c2V{ \cos { 2 \theta_{D^*}}}

\def\s2l{ \sin {2 \theta_{l}}}

\def\c2l{ \cos { 2 \theta_{l}}}

\def\roughly#1{\mathrel{\raise.3ex\hbox
{$#1$\kern-.75em\lower1ex\hbox{$\sim$}}}}

\def\barp{{\raise.35ex\hbox
{${\sss (}$}}---{\raise.35ex\hbox{${\sss )}$}}}
\def\bdbarp{\hbox{$B_d$\kern-1.4em\raise1.4ex\hbox{\barp}}}
\def\bbarp{\hbox{$B$\kern-1.1em\raise1.4ex\hbox{\barp}}}
\def\sss{\scriptscriptstyle}

  \def\rr2{{1\over\sqrt{2}}}
\def\nn{\nonumber}
\def\sss{\scriptscriptstyle}
\def\t3{{1\over\sqrt{3}}}
\def\s6{{1\over\sqrt{6}}}
\def\rr2{{1\over\sqrt{2}}}
\def\nl{\nonumber\\}
\def \({\left(}
\def \){\right)}
\def \[{\left[}
\def \]{\right]}
\def \l|{\left|}
\def \r|{\right|}

\def \nn{\nonumber}
\def \nl{\nn \\}

\def \be{\beta}

\def \s{\sqrt{2}}

\def \lc{\Lambda_c^+}

\def \RD{R_{D}^{\tau/\ell}}
\def \RDMu{R_{D}^{\tau/\mu}}
\def \RDstar{R_{D^*}^{\tau/\ell}}

\def\Rlc{{R(\Lambda_c^+)}^{\tau/\ell}}
\def\RlcMu{{R(\Lambda_c^+)}^{\tau/\mu}}

\def \cL {{\cal L}}

\def \s{\sqrt{2}}

\def \be{\beta}

\def \cL{{\cal L}}

\def \({\left(}
\def \){\right)}
\def \[{\left[}
\def \]{\right]}

\newcommand*{\rom}[1]{\expandafter\@slowromancap\romannumeral #1@}

\def\cL{{\cal L}}

\def\beq{\begin{equation}}
\def\eeq{\end{equation}}
\def\beqa{\begin{eqnarray}}
\def\eeqa{\end{eqnarray}}

\makeatletter

\begin{document} 
\title{Hint of a new scalar interaction in LHCb data?}

\author{Alakabha Datta}
\email{datta@phy.olemiss.edu}
\affiliation{Department of Physics and Astronomy,
108 Lewis Hall, University of Mississippi, Oxford, MS 38677-1848, USA.}

\author{Danny Marfatia}
\email{dmarf8@hawaii.edu}
\affiliation{Department of Physics and Astronomy, University of Hawaii at Manoa, 2505 Correa Rd., Honolulu, HI 96822, USA.}

\author{Lopamudra Mukherjee}
\email{lopamudra@nankai.edu.cn}
\affiliation{Department of Physics, Indian Institute of Technology Guwahati, Assam 781039, India.}
\affiliation{School of Physics, Nankai University, Tianjin 300071, China.}
\begin{abstract}
We explain recent LHCb measurements of the lepton universality ratios, 
$R_{D^{(*)}}^{\tau/\ell}\equiv \frac{\mathcal{B}(\bar B \to D^{(*)+} \tau^- \bar\nu_\tau)} {\mathcal{B}(\bar B \to D^{(*)+}\ell^- \bar\nu_\ell)}$
and ${R(\Lambda_c^+)}^{\tau/\ell} \equiv \frac{\mathcal{B}(\Lambda_b \to \Lambda_c^+ \tau^- \bar{\nu}_{\tau})}{\mathcal{B}(\Lambda_b \to \Lambda_c^+ \ell^- \bar{\nu}_{\ell})}$ with $\ell=\mu$, via new physics that affects $R_D^{\tau/\ell}$ and $R(\Lambda_c^+)^{\tau/\ell}$ but not $R_{D^*}^{\tau/\ell}$. The scalar operator in the effective theory for new physics is indicated. 
We find that the forward-backward asymmetry and $\tau$ polarization in $\bar{B} \to D^+ \tau^{-} \bar{\nu}_{\tau}$ and $\Lambda_b \to \Lambda_c^+ \tau^- \bar{\nu}_{\tau}$ decays are significantly affected by the scalar interaction.
 We construct a simple two Higgs doublet model as a realization of our scenario and consider lepton universality in semileptonic charm and top decays, radiative $B$ decay, $B$-mixing, and $Z \to b \bar b$.
 \end{abstract}

\maketitle

{\bf{Introduction.}}
\label{sec:introduction}
A major part of particle physics research is focused on searching for physics beyond the standard model (SM). A key property of the SM gauge interactions is that they are lepton flavor universal. Evidence for violation of this property would be a clear sign of new physics (NP).  Many measurements are sensitive to the violation of flavor universality. In such searches, the second and third generation quarks and leptons are 
special because they are comparatively heavier, making their interactions relatively more sensitive to NP in some scenarios. 
As an example, in certain versions of the two Higgs doublet model (2HDM), the couplings of the new Higgs bosons are proportional to the fermion masses and so lepton universality effects are more pronounced for the heavier generations. Moreover, constraints on NP involving third generation leptons and quarks are somewhat weaker, allowing for larger NP effects. 
In this Letter, we focus on measurements in the $b$ quark system which show hints of lepton flavor universality violation in certain decays.

The charged-current decays,  $B \to D^{(\ast)} \tau  \nu_\tau$,  have been observed by the BaBar, Belle  and the LHCb  experiments. Measurements of
  $R_{D^{(*)}}^{\tau/\ell} \equiv \frac{\cB(\bar{B}
\to D^{(*)} \tau^{-} {\bar\nu}_\tau)}{\cB(\bar{B} \to D^{(*)} \ell^{-}
{\bar\nu}_\ell)}$ ($\ell = e,\mu$)~\cite{BaBar:2012obs, BaBar:2013mob, Belle:2015qfa, Belle:2016ure, Belle:2016dyj,  Belle:2017ilt, Belle:2019rba, LHCb:2023zxo, LHCb:2023cjr} and $R_{J/\psi} \equiv \frac{\cB(B_c^+ \to J/\psi\tau^+\nu_\tau)}{\cB(B_c^+ \to J/\psi\mu^+\nu_\mu)}$~\cite{LHCb:2017vlu} are discrepant with predictions of
the SM and provide a hint of lepton universality violation in $\bctaunutau$ decay. The combined BaBar, Belle and LHCb data show $\sim 2\sigma$ discrepancies in both $R_D^{\tau/\mu}$ and $R_{D^*}^{\tau/\mu}$. However, BaBar \cite{BaBar:2012obs,BaBar:2013mob} and Belle \cite{Belle:2015qfa, Belle:2016ure, Belle:2016dyj, Belle:2017ilt, Belle:2019rba}
found discrepancies in either both modes or in neither mode. Only LHCb \cite{LHCb:2023zxo, LHCb:2023cjr} finds a discrepancy in one mode and not the other. The ratios of branching fractions, $\RD$ and $\RDstar$, have an advantage over the absolute branching fraction measurements of $ B \to D^{(\ast)} \tau  \nu_\tau$, as these ratios are  relatively less sensitive to the form factor and systematic uncertainties,
such as those from the experimental efficiency and the value of $|V_{cb}|$, which cancel in the ratios.
Nevertheless, $R_{J/\psi}$ does not provide as good a test of lepton universality since the experimental measurement is currently less precise. 

Recently, the LHCb collaboration presented its first simultaneous measurement of $R_{D^{(*)}}^{\tau/\mu}$ with the $\tau$ reconstructed using the semileptonic decay $\tau^+ \to \mu^+ \nu \bar\nu$~\cite{LHCb:2023zxo} as listed in Table~\ref{tab:obs_meas}. 
Moreover, a new result for $R_{D^*}^{\tau/\ell}$ with hadronic $\tau$ decays was presented  by the LHCb collaboration, which when combined with the Run 1 result gives a value of $R_{D^*}^{\tau/\ell}$~\cite{LHCb:2023cjr} which is in excellent agreement with the SM prediction; see Table~\ref{tab:obs_meas}.

\begin{table}[t]
\renewcommand{\arraystretch}{1.5}
\resizebox{\columnwidth}{!}{
\begin{tabular}{|c|c|c|c|} \hline
Observable & SM expectation & LHCb measurement(s) & Deviation\\
\hline
$R_{D}^{\tau/\ell}$ & $0.298 \pm 0.004$  \cite{HFLAV-WA-2021} & $0.441 \pm 0.060 \pm 0.066$ \cite{LHCb:2023zxo} &  $1.6\sigma$\\\hline
\multirow{2}{*}{$R_{D^*}^{\tau/\ell}$} & \multirow{2}{*}{$0.254 \pm 0.005$ \cite{HFLAV-WA-2021}} & $0.281 \pm 0.018 \pm 0.024$ \cite{LHCb:2023zxo} & $0.89\sigma$ \\
& & $0.257 \pm 0.012 \pm 0.014 \pm 0.012$ \cite{LHCb:2023cjr} &  $0.13\sigma$ \\\hline
$R(\lc)^{\tau/\mu} $ & $0.333 \pm 0.010$ \cite{Detmold:2015aaa} & $0.242 \pm 0.026 \pm 0.059$ \cite{LHCb:2022piu} & $-1.4\sigma$ \\ \hline
\end{tabular}
}
\caption{LHCb measured values of lepton flavor universality violating observables and their deviations from SM expectations.
LHCb is the first experiment to find a discrepancy in 
$R_{D}^{\tau/\ell}$ (although at only $\sim 90$\%~C.L.), but not $R_{D^*}^{\tau/\ell}$. }
\label{tab:obs_meas}
\end{table}

Based on these recent developments, we entertain the possibility that NP affects
$R_{D}^{\tau/\ell}$ but not $R_{D^*}^{\tau/\ell}$. 
To establish what kind of physics would yield this scenario, it is useful to consider an effective theory description of the NP. Only a scalar hadronic current
in the four fermion interaction works because $D^*$ is a vector meson. In principle, this has implications for other decays that proceed through the same underlying $\bctaunutau$ transition such as
$\lbt$, $B_c^+ \to J/\psi\tau^+\nu_\tau$ and $ B_c \to \tau^- \nu_\tau$. However, the latter two decays are unaffected.
It is interesting that LHCb has reported a measurement of
$R(\lc)^{\tau/\ell}  =  \frac{{\cal B}( \lbt)}{{\cal B}(\lbl)}$ with $\ell=\mu$~\cite{LHCb:2022piu}\footnote{Reference~\cite{Bernlochner:2022hyz} finds that the measured value of $R(\lc)^{\tau/\mu}$ is closer to the SM value if
 the  $\Lambda_b \to \Lambda_c \tau^- \nu_\tau$ rate is normalized to the SM prediction for $\Lambda_b \to \Lambda_c \mu^- \nu_\mu$ instead of normalizing to old experimental measurements.
 By doing so, Ref.~\cite{Bernlochner:2022hyz} obtains $R(\lc)^{\tau/\mu} = |0.04/V_{cb}|^2 (0.285 \pm 0.073)$ which is consistent with the SM prediction within its uncertainty. However, we adopt the LHCb measured value in Table~\ref{tab:obs_meas}. 
} which shows a $1.4\sigma$ deficit relative to the SM prediction.
Note that Refs.~\cite{Bernlochner:2018kxh, Bernlochner:2018bfn} find
 $R(\lc)^{\tau/\mu}_{\rm {SM}} = 0.324 \pm 0.004$ using
  the heavy quark expansion, lattice results and experimental input from
 $\lbl$ decays assuming these decays are described by the SM. The SM expectation
 in Table~\ref{tab:obs_meas} agrees with these analyses within  uncertainties. 
 In this work, we study how the scalar interaction impacts $R(\lc)^{\tau/\mu}$
and make predictions for several observables for semileptonic $B$ and $\Lambda_b$  decays that can be used to test the scenario. Global fits to $b \to c \tau \nu$ data in an effective theory framework have been performed in Refs.~\cite{Murgui:2019czp, Iguro:2022yzr,Ray:2023xjn} but these do not consider
an enhanced $R_D^{\tau/\ell}$ and a SM $R_{D^{*}}^{\tau/\ell}$ with just a scalar operator.

The next step is to construct a simple partial ultraviolet completion of our scenario. Typically, to explain the $R_{D^{(*)}}^{\tau/\ell}$ measurements, models with extra gauge bosons, scalars and leptoquarks are considered. The scalar nature of the new interaction rules out extra gauge bosons, and requires unnatural correlations between couplings of different types of leptoquarks~\cite{Dumont:2016xpj}. The only natural scenario is one with extra scalars such as the two Higgs doublet model. We will explore the phenomenological implications of the 2HDM that only generates the new scalar operator for semileptonic $B$ decays.

%

%

{\bf{Effective Hamiltonian.}}
\label{sec:formalism}
The effective Hamiltonian that describes NP in the quark-level transition, $b\to c\tau^-\bar{\nu}_\tau$, can be written at the $m_b$ scale in the form~\cite{Chen:2005gr,Bhattacharya:2011qm},
\bea
\label{eq1:Lag}
 {\cal{H}}_{eff} &=&  \frac{G_F V_{cb}}{\sqrt{2}}\Big\{
\Big[\bar{c} \gamma_\mu (1-\gamma_5) b  + g_L \bar{c} \gamma_\mu (1-\gamma_5)  b \nl &+& g_R \bar{c} \gamma_\mu (1+\gamma_5) b\Big] \bar{\tau} \gamma^\mu(1-\gamma_5) \nu_{\tau} \nl &+&  \Big[g_S\bar{c}  b   + g_P \bar{c} \gamma_5 b\Big] \bar{\tau} (1-\gamma_5)\nu_{\tau} \nl & +& \Big[g_T\bar{c}\sigma^{\mu \nu}(1-\gamma_5)b\Big]\bar{\tau}\sigma_{\mu \nu}(1-\gamma_5)\nu_{\tau} + {{\rm h.c.}} \Big\}, \label{eq:Heff}
\eea 
where  $G_F$ is the Fermi constant, $V_{cb}$ is the Cabibbo-Kobayashi-Maskawa (CKM) matrix element, and $\sigma_{\mu \nu} = i[\gamma_\mu, \gamma_\nu]/2$.{\footnote{If the effective interaction is written at the cut-off scale $\Lambda$ then renormalization group evolution to the $m_b$ scale will generate new operators
which have been discussed in Refs.~\cite{Feruglio:2016gvd, Feruglio:2017rjo}. These new contributions can strongly constrain models but depending on the model,
 there may be cancellations between various terms.}

The SM Hamiltonian corresponds to $g_L = g_R = g_S = g_P = g_T = 0$.
In Eq.~\eqref{eq:Heff}, we have taken the neutrinos to be always left chiral. In general, with NP the neutrino associated with the $\tau$ lepton does not have to carry the same flavor, but we will not consider this possibility.

{\bf{Observables.}}
\label{sec:Obs}
The angular distributions of $\BDtaunu$ and $\lbt$ can be used to measure several observables such as branching ratios, angular variables and final state lepton polarizations. The differential distributions in the various kinematic and angular variables are expressed in terms of helicity amplitudes which depend on the Wilson coefficients of the new physics operators. 
In the following, we present the definitions of these observables for $\BDtaunu$ and $\lbt$.

\paragraph{$\BDtaunu$.}
For our analysis, in addition to $\RD$, the observables of interest are 
(1)~Forward-backward asymmetry:
\beq 
\mathcal{A}_{FB,\tau}^D(q^2) = \frac{1}{d\Gamma/dq^2} \[\int_0^{1}- \int_{-1}^{0}\]\frac{d^2\Gamma}{dq^2d\cos \theta_\tau}d\cos \theta_\tau\,,
\eeq 
where $\theta_\tau$ is the angle between the $D$ and $\tau$ in the centre of mass frame of the dilepton, and 
(2)~Lepton polarization asymmetry:
\beq
P_\tau^D(q^2) = \frac{1}{d\Gamma/dq^2}\[\frac{d\Gamma^{\lambda_\tau = +1/2}}{dq^2}-\frac{d\Gamma^{\lambda_\tau = -1/2}}{dq^2}\]\,,
\eeq 
where $\lambda_\tau = \pm 1/2$ denotes the helicity of the  $\tau$ lepton. The total rate $\Gamma = \Gamma^{\lambda_\tau = +1/2} + \Gamma^{\lambda_\tau = -1/2}$. 

\paragraph{$\lbt$.}
\label{sec:lbt}
Besides $\Rlc$, observables that can be used to test for new physics are 
(1) $\mathcal{A}_{FB,\tau}^{\Lambda}(q^2)$ defined in analogy with $\mathcal{A}_{FB,\tau}^D(q^2)$,
(2) $P_\tau^\Lambda(q^2)$ defined in analogy with $P_\tau^D(q^2)$,
(3) the $\pi/3$ asymmetry defined as~\cite{Becirevic:2022bev}
\beq
\footnotesize
\mathcal{A}_{\pi/3}^\Lambda (q^2) = {1 \over d\Gamma/dq^2} \[\int_0^{\pi/3} +\int_{2\pi/3}^\pi -\int_{\pi/3}^{2\pi/3}\] {d^2\Gamma \over dq^2 d\cos \theta_\tau} \sin \theta_\tau d\theta_\tau\,,
\eeq 
and (4) the asymmetry in the azimuthal angle distribution~\cite{Becirevic:2022bev},
\beq
D_4^\Lambda(q^2) = {1 \over d\Gamma/dq^2}\[\int_0^\pi - \int_\pi^{2\pi}\]{d^2\Gamma \over dq^2d\phi}d\phi\,,
\eeq 
where $\phi$ is the angle between the decay planes of $\Lambda_c^+ \to \Lambda^0\pi^+$ and $W^- \to \tau^- \bar\nu_\tau$. Since it is a CP-violating asymmetry, it is nonzero only for complex Wilson coefficients. $D_4$ is proportional to the asymmetry in the $\Lambda_c \to \Lambda \pi$ decay parametrized by $\alpha$. For the NP predictions we use $\alpha = -0.84$ \cite{Becirevic:2022bev}.

{\bf{Data analysis.}}
\label{sec:Fits} 
We set $g_L = g_R = g_P = g_T = 0$ and fit the real and imaginary parts of $g_S$ to the measured values of $R_D^{\tau/\mu}$ and $R(\Lambda_c^+)^{\tau/\mu}$ in Table~\ref{tab:obs_meas}.
We utilize the $B \to D\ell \nu$ form factors, at nonzero recoil obtained by the HPQCD~\cite{Na:2015kha} and MILC~\cite{MILC:2015uhg} to extract the form factor parameters using the BGL parameterization~\cite{PhysRevD.56.6895}. For $\Lambda_b \to \Lambda_c$ decay, we use the form factor fit results of Ref.~\cite{Datta:2017aue} to the lattice QCD results in Ref.~\cite{Detmold:2015aaa} using BCL z-expansions~\cite{Bourrely:2008za}. Our SM results are based only on these lattice fit results without any additional experimental input. We set the decay scale $\mu = m_b = 4.18$~GeV,
$m_b(m_b) = 4.18$~GeV and $m_c(m_b) = 0.92$~GeV. 

\begin{table}[t]
    \centering
    \begin{tabular}{|c|c|c|}
    \hline 
    \multicolumn{2}{|c|}{Form factor parameters}\\
    \hline
    $B\to D$ & $\Lambda_b \to \Lambda_c$ \\
    \hline \hline
    $a_0^{f_+} = 0.0156$& $a_0^{F_+} = 0.8313$, $a_1^{F_+} = -4.3562$  \\
    $a_1^{f_+} = -0.0420$& $a_0^{F_0} = 0.7314$, $a_1^{F_0} = -5.2560$\\
    $a_2^{f_+} = -0.1097$& $a_0^{F_\perp} = 1.1036$, $a_1^{F_\perp} = -5.5691$\\
    $a_0^{f_0} = 0.0779$& $a_0^{G_+} = 0.6932$, $a_1^{G_+} = -4.0740$\\
    $a_1^{f_0} = -0.1972$& $a_0^{G_0} = 0.7254$, $a_1^{G_0} = -4.6979$\\
    & $a_1^{G_\perp} = -4.1020$\\
    \hline \hline
    \end{tabular}
    \caption{Best fit form factor parameters which are identical for both real $g_S$ and complex $g_S$.}
    \label{tab:fitrslt}
\end{table}

We minimize the $\chi^2$ function defined as $\chi^2 = (\mathcal{E} - \mathcal{O})^T \mathcal{C}^{-1} (\mathcal{E} - \mathcal{O})+\chi^2_{nuis}$, where $\mathcal{E}$ denotes the vector of measurements, $\mathcal{O}$ is the corresponding vector of predicted values, and $\mathcal{C}$ is the covariance matrix of the measurements. 
Here, $\chi^2_{nuis} = (\mathcal{N} - \mathcal{P})^T \mathcal{V}^{-1} (\mathcal{N} - \mathcal{P})$ where $\mathcal{N}$ is the vector of form factor values, $\mathcal{P}$ is the corresponding vector of the fit parameters and $\mathcal{V}$ is the covariance matrix
corresponding to $\mathcal{N}$.
We treat the form factor parameters as nuisance parameters and marginalize over them.
The $x$-$\sigma$ uncertainty on a fit parameter is obtained by requiring $\Delta \chi^2 = \chi^2 - \chi^2_{min} \leq x$ after marginalizing over all other parameters. For the real $g_S$ scenario, we obtain two best fit points at $g_S = 0.171$ and $g_S = -1.60$, both with $\chi^2_{min} = 1.6$ and $pull = \sqrt{\chi^2_{\rm {SM}}-\chi^2_{min}} = 1.48$ with respect to the SM. The $1\sigma$ allowed range for the real $g_S$ scenario is 
$[-1.77,-1.47]\oplus [0.039,0.314]$.
The complex $g_S$ scenario does not lead to a lower $\chi^2_{min}$ than the real $g_S$ scenario, and the imaginary component is only mildly constrained. 
We present the values of the best fit form factor parameters, applicable to both scenarios, in Table~\ref{tab:fitrslt}. 
In the top panel of Fig.~\ref{fig:RD-Rlamc}, the black dashed ellipse is the locus of best fit points with $\chi^2_{min} = 1.6$.
The $1\sigma$ allowed region for the real and imaginary components of $g_S$ is also shown. In the bottom panel, we compare the theoretical predictions for the fit observables with the SM expectation and the LHCb measured values. The black and blue points mark the SM and LHCb measurements, respectively.  The pink scatter points show the 1$\sigma$ C.L. region, corresponding to $\Delta \chi^2 \leq 1$, for the real $g_S$ scenario. Similarly, for the complex $g_S$ scenario, we plot the 1$\sigma$ C.L. regions in green by selecting $g_S$ values for which $\Delta \chi^2 \leq 2.30$. 
We select five best fit points from the dashed ellipse in Fig.~\ref{fig:RD-Rlamc}, and provide the corresponding predictions for several observables related to $\BDtaunu$ and $\lbt$ in Table~\ref{tab:predictions}.
The SM expectations are also listed.

\begin{figure}[t]
    \centering
    \includegraphics[scale=0.5]{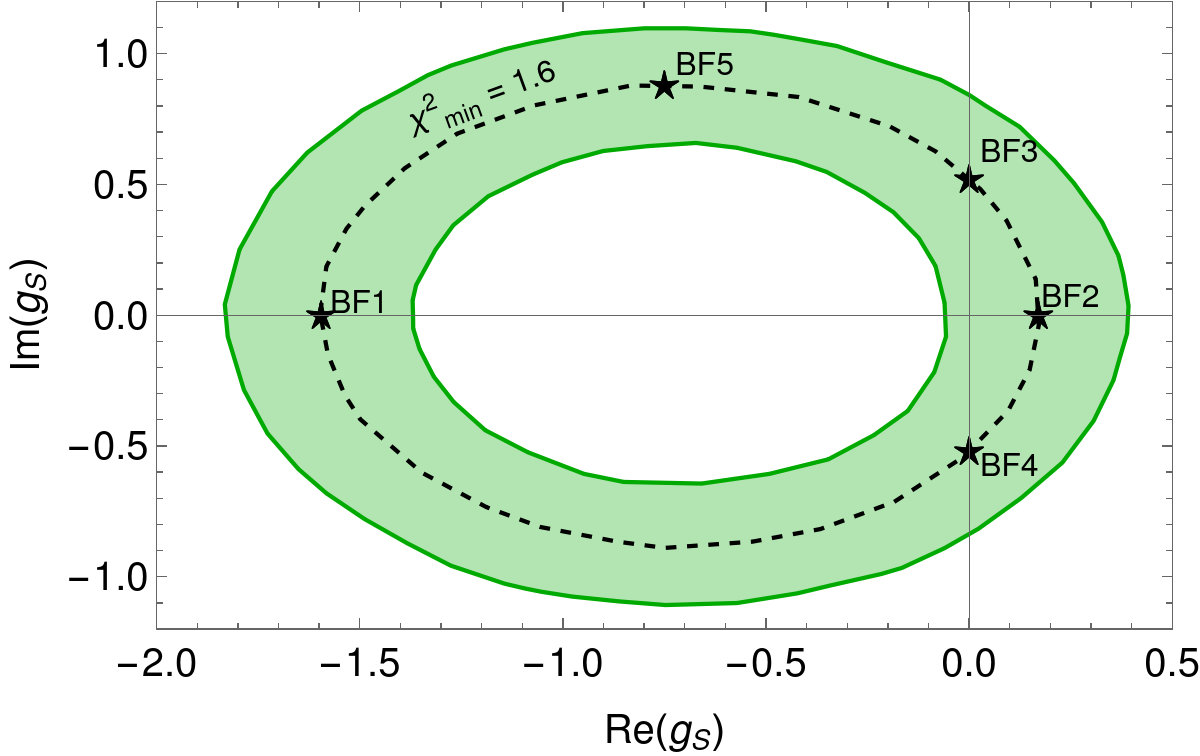}
    \includegraphics[scale=0.5]{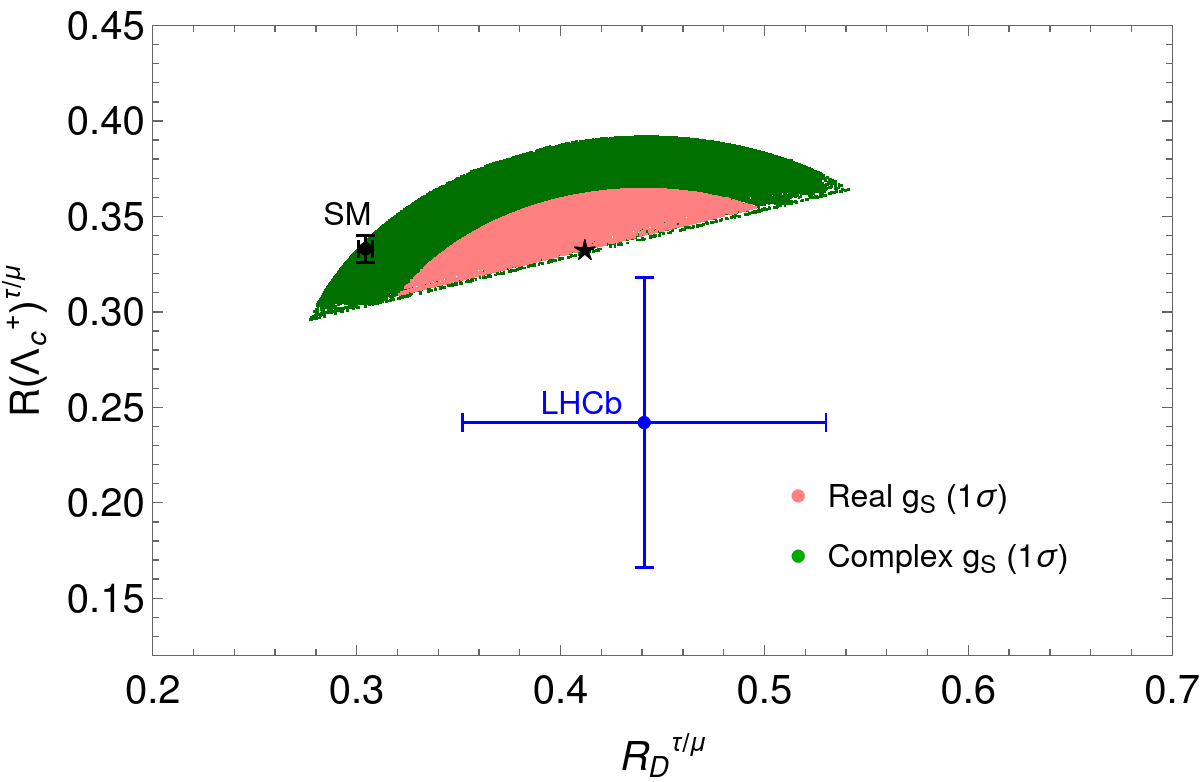}
    \caption{Top: The $1\sigma$ allowed parameter space for complex $g_S$. The locus of best fit points defined by $\chi^2_{min}=1.6$ is shown by the black dashed ellipse. 
    The stars mark the five best fit points in Table~\ref{tab:predictions}.
    Bottom: SM and NP predictions for two observables. The star marks the best fit prediction, which is the same for all best fit points.  The values measured by LHCb are shown for comparison. The ellipse is cut in half because the new scalar interaction cannot simultaneously enhance $\RD$ and suppress $\Rlc$.}
    \label{fig:RD-Rlamc}
\end{figure}

\begin{table*}[t]
    \centering
    \renewcommand{\arraystretch}{1.5}
    \resizebox{\textwidth}{!}{
    \begin{tabular}{c|c|c|c|c|c|c|c}
    \hline \hline
    \multirow{4}{*}{Observable}  &  \multicolumn{7}{|c}{Prediction}  \\
    \cline{2-8}
    & \multirow{3}{*}{SM} & \multicolumn{6}{|c}{New physics}\\
    \cline{3-8}
    & & \multirow{2}{*}{Real $g_S$: $1\sigma$ range} & \multirow{2}{*}{Complex $g_S$: $1\sigma$ range} & BF1 & BF2 & BF3/BF4 & BF5 \\
    &&& & $g_S = -1.60 + 0i$ & $g_S = 0.171 + 0i$ & $g_S = 0 \pm 0.52 i$ & $g_S = -0.75 + 0.88i$ \\
    \hline \hline
    $\RDMu$ & $0.304(3)$ & $[0.324, 0.485]$& [0.277, 0.541] & $0.412$ & $0.412$ & $0.411$ & $0.410$ \\  
    $\RlcMu$ & $0.333(7)$ &  $[0.309, 0.364]$ & [0.296, 0.392] & $0.333$ & $0.333$ & $0.333$ & $0.332$ \\
    $\langle P_\tau^D \rangle$ & $0.324(3)$ & [0.355, 0.590]& $[0.252, 0.623]$ & $0.492$ & $0.492$ & $0.491$ & $0.490$\\
    $\langle P_\tau^\Lambda \rangle$ & -0.308(6) &  [-0.283, -0.113] & [-0.350, -0.078] & $-0.259$ & $-0.258$ & $-0.259$ & $-0.260$\\
    $\langle A_{FB,\tau}^{D}\rangle$ & $0.3596(4)$ &  $[0.0416,0.1149]\oplus[0.2589,0.3446]$& $[0.0331, 0.3811]$ & $0.0743$ & $0.297$ & $0.276$ & $0.182$ \\
    $\langle A_{FB,\tau}^{\Lambda}\rangle$ & $0.024(4)$ & [-0.170,-0.145]$\oplus$[0.027, 0.061] & [-0.170, -0.068] & $-0.165$ & $0.028$ & $0.0097$ & $-0.073$\\
    $\langle \mathcal{A}_{\pi/3}^\Lambda \rangle$ & -0.0264(6) &  [-0.028, -0.022] & [-0.029,-0.021] & $-0.0258$ & $-0.0258$ & $-0.0259$& $-0.0259$\\
    $\langle D_4^\Lambda\rangle$ &$0$ & $-$ & [-0.089,0.089] & $0$ & $0$ & $\pm 0.044$ & $0.075$\\
    \hline \hline
    \end{tabular}
    }
    \caption{SM and NP predictions for various observables. For NP, we list the predictions for five best fit points. The SM expectations are based on the lattice form factor results but no other experimental input.}
    \label{tab:predictions}
\end{table*}

It is evident from Fig.~\ref{fig:RD-Rlamc} and Table~\ref{tab:predictions}, that the deficit in $\tau$ production in the $\Lambda_b \to \Lambda_c$ decays cannot be accommodated by the scalar interaction. In fact, the global minima are at values of $g_S$ for which $\RlcMu = R(\lc)^{\tau/\mu}_{\rm {SM}}$. This is in line with the sum rule of Refs.~\cite{Iguro:2022yzr, Fedele:2022iib} that relates $R_{D^{(*)}}^{\tau/\ell}$ and $R(\lc)^{\tau/\ell}$.  However, note that the sum rule does not allow for variations in the form factors which is important for interpreting new physics. 
In Fig.~\ref{fig:Rvsgs}, we show how $\RDMu$ and $\RlcMu$ are affected by the choice of form factor. Results for form factors based solely on lattice fit values are shown by the black curves, and those for our best fit form factors are shown in red.  

\begin{figure}[t]
    \centering
    \includegraphics[scale=0.5]{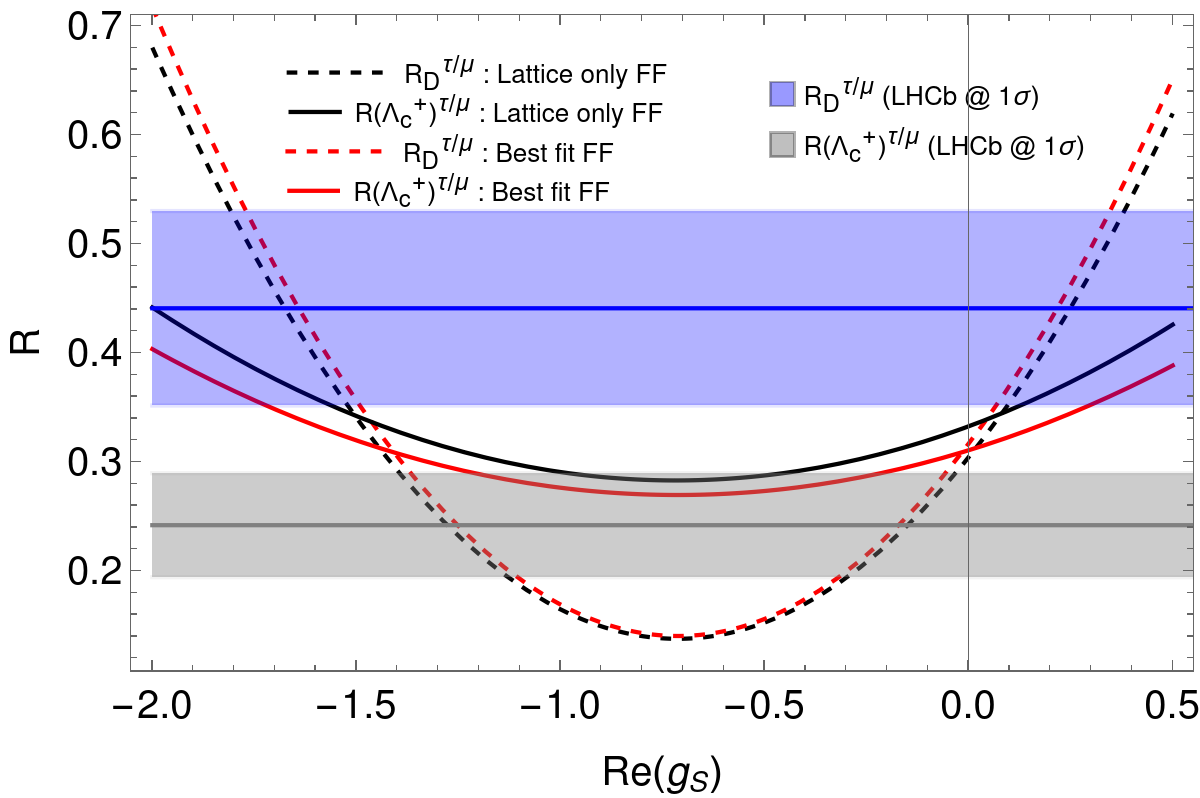}
    \caption{$\RDMu$ (dashed curves) and $\RlcMu$ (solid curves) with Im$(g_S)=0$ for the form factors obtained from lattice fits (black) and for best fit form factors  in Table~\ref{tab:fitrslt} (red). The blue and gray bands represent the $1\sigma$ ranges of the respective LHCb measurements.}
    \label{fig:Rvsgs}
\end{figure}

As indicated in Table~\ref{tab:predictions}, the best fit scenarios yield SM-like $\RlcMu$ and $\langle A_{\pi/3}^\Lambda \rangle$. For all the best fit points we observe an enhancement in $\langle P_\tau^D \rangle$ and a relative suppression in $|\langle P_\tau^\Lambda \rangle|$ with respect to the SM, making these observables good discriminators of scenarios with new scalar physics. 
Note that the sign of $\langle A_{FB,\tau}^\Lambda\rangle$ depends on the value of $g_S$. Of our observables, $\langle D_4^\Lambda\rangle$ is the only truly CP violating angular asymmetry and is sensitive to Im$(g_S)$. Depending on the sign of the imaginary component, we obtain either a positive or negative shift from the SM expectation of zero. Therefore, measurements of these angular observables will be important to determine whether the new scalar current is real or complex in nature and to distinguish between the best fit scenarios.

{\bf{Model.}}
\label{sec:UVC}
A widely studied class of models is an extension of the
SM with  additional scalar SU(2) doublets, the simplest of which are two Higgs doublet models~\cite{Lee:1973iz,Branco:2011iw,Davidson:2005cw}.  Generally, when quarks couple to more than one scalar doublet, flavor changing neutral currents (FCNCs) are generated since the diagonalization of the up-type and down-type mass matrices does not automatically ensure that the couplings to each scalar doublet are also diagonal.  We focus on the Yukawa couplings of the Higgs sector and
consider the most general Yukawa Lagrangian of the form,
\bea
{\cal L}^{(q)}_{Y}& = & \eta^{U}_{ij} \bar Q_{i,L} \tilde\phi_1 U_{j,R} +
\eta^D_{ij} \bar Q_{i,L}\phi_1 D_{j,R} \nonumber\\ && + \xi^{U}_{ij} \bar Q_{i,L}\tilde\phi_2 U_{j,R}
+\xi^D_{ij}\bar Q_{i,L} \phi_2 D_{j,R} \,+\, {\rm h.c.}\,,  \nonumber\\
{\cal L}^{(l)}_{Y} & = & 
\eta^E_{ij} \bar L_{i,L}\phi_1 E_{j,R} + \xi^{E}_{ij} \bar L_{i,L}\phi_2 E_{j,R} \,+\, {\rm h.c.}\,, 
\label{2HDM}
\eea

\noindent where $Q_{i,L}$, $U_{j,R}$, and $D_{j,R}$ are the quark flavor eigenstates, $\phi_i$ with $i=1,2$, are the two scalar doublets, $\tilde\phi_i = i\sigma_2 \phi_i^*$,  and $\eta^{U,D,E}_{ij}$ and $\xi_{ij}^{U,D,E}$ are the nondiagonal matrices of the Yukawa couplings.  
With no discrete symmetry imposed, both up-type and down-type quarks can have FCNC couplings \cite{Pakvasa:1977in}. 

In the notation and basis of Ref.~\cite{Atwood:1996vj}, the  general charged Higgs boson $H^{\pm}$ interactions in the fermion mass basis are given by
%
\beq
\label{eq:Lag}
\footnotesize
\cL^{H^+}= -\bar u (\!V\cdot
\hat\xi^{D} P_R -{\hat\xi^{U}}\cdot 
V P_L )d \, H^+ 
 -  \bar \nu \, \hat \xi^{E}P_R \,e\, H^+
 + \text{h.c.}\,,
\eeq
where $\hat \xi^{f}$ ($f= u, d, e$) are $3\times 3$ real NP Yukawa matrices,
$V$ is the CKM matrix, and $P_{L/R}\equiv (1\mp \gamma_5)/2$ are the chirality projectors.
To simplify the discussion we neglect neutrino masses and mixing.
Defining $V_{L,R}^{U,D}$ to be
the rotation matrices acting on the up- and down-type quarks, with left or right chirality respectively, 
the neutral flavor changing couplings are
%
$\hat\xi^{U,D}=(V_L^{U,D})^{\dagger}\  \xi^{U,D}\ V_R^{U,D}$.
In our calculations we use the flavor ansatz
 \bea
 \label{UP}
 \hat \xi^{U}& = & - V \hat \xi^D V^\dagger\,,
 \label{fmodel}
 \eea
 where $V= V_L^{ U \dagger} V_L^D$ is the CKM matrix for the left handed charged current interactions.
The structure in Eq.~(\ref{fmodel}) ensures a scalar hadronic current from
Eq.~(\ref{eq:Lag}) and no pseudoscalar hadronic coupling at tree level so that the charged Higgs interactions become
\mbox{$\cL^{H^+}= -\bar u V\cdot
\hat\xi^{D}   d \, H^+ 
 -  \bar \nu \, \hat \xi^{E}P_R \,e\, H^+
 + \text{h.c.}$.}
 Since the matrix element $\langle{D^*}| \bar{c}b| B\rangle$ vanishes there is no contribution to $\RDstar$ in this model. Note that a pseudoscalar current
 is also severely constrained by $B_c \to \tau^- \nu_\tau$ decay~\cite{Alonso:2016oyd}.

 It is interesting to speculate on a possible origin of the flavor ansatz  in Eq.~(\ref{fmodel}).
Requiring the $\phi_2$ Yukawa interaction in Eq.~(\ref{2HDM})
to be invariant under the discrete transformations $U_R \leftrightarrow D_R$ and $ \tilde \phi_2 \leftrightarrow - \phi_2$, yields  $\xi^U=- \xi^D$.
This leads to
$ \hat \xi^{U} =  - V \hat \xi^D V^{\prime \dagger}$,
where $V^\prime= V_R^{ U \dagger} V_R^D$ is the mixing matrix for the right handed charged current interactions.
 If $V^\prime = V$, we recover the ansatz in  Eq.~(\ref{fmodel}).
 We point out that  various versions of 2HDM with natural flavor conservation have difficulty in explaining $\RD$ and $\RDstar$ while being consistent with other constraints~\cite{Fajfer:2012jt}.
 A more general flavor structure must be resorted to~\cite{Crivellin:2012ye,Celis:2012dk}
 but a joint explanation of $R_{D^(*)}^{\tau/\ell}$ generally remains problematic.

To explain the LHCb data with a minimal set of NP parameters,
we make the  further ansatz that the NP Yukawa matrices have the following simple structures:
\begin{align}\label{eq:texture}
 \hat \xi^{D} = \lambda_{q} 	\begin{pmatrix}
	1 & 0 & 0\\0 &  1  &0 \\0 & 0 & 1\\
\end{pmatrix},\quad \hat \xi^{E} = 
	\begin{pmatrix}
	0 & 0 & 0\\0 & 0 & 0\\0 & 0 & \lambda_{\tau \tau}\\
\end{pmatrix}\,.
\end{align}
The ansatz for the leptonic mixing matrix is similar to the  Aligned 2HDM~\cite{Pich:2009sp}
in which the alignment condition is applied to the lepton sector. Then, the charged Higgs interaction is
$\cL^{H^+}= 
 -\frac{ \sqrt{2}}{v}\bar \nu \, \zeta_l M_lP_R \,e\, H^+
 + \text{h.c.}$,
where $M_l$ are the charged lepton diagonal mass matrices and $v$ is the Higgs vacuum expectation value.
The $\zeta_{l}$ are  free parameters that connect the Yukawa couplings of the first Higgs doublet to the Yukawa couplings of the second Higgs doublet.  Under the approximation that the $ \mu$ and e masses are negligible, we recover our leptonic current.

Using the textures in Eq.~\eqref{eq:texture} we obtain
 \begin{align}
 \label{eq:Lag2}
\cL^{H^+} = -\lambda_{q} \left [ V_{ij}\bar u_i d_j  \right] H^+ 
 -  \lambda_{\tau \tau}\bar \nu_{\tau} \, P_R \,\tau\, H^+
 + \text{h.c.}\,.
\end{align}

It is clear from Eq.~\eqref{eq:Lag2} that new physics effects are present in  all semileptonic transitions, but since $H^+$ couples to the $\tau$ lepton, only the charmed, bottom and top semileptonic decays will be affected due to kinematics.
The effective Hamiltonian for $b\to c\tau^-\bar{\nu}_\tau$ is then (with Hermitian conjugate terms implicit),
\beq\label{eq:effLag_b2ctaunu}
	{\cal{H}}_{eff} =  \frac{\lambda_{bc}\lambda_{\tau \tau}^\ast}{2 M_{H_{\pm}}^2} \bar c  b  \ \bar{\tau} (1-\gamma_5)\nu_{\tau}   
  =  \frac{G_F V_{cb}}{\sqrt{2}} g_S \bar c b  \ \bar{\tau} (1-\gamma_5)\nu_{\tau}\,,
 \eeq
 where $\lambda_{bc}= \lambda_q V_{cb}$
and $g_S$ at the scale $\mu \sim  m_b$ is given by
\begin{align}\label{eq:CSL}
	g_S = \frac{\lambda_{q}\lambda_{\tau \tau}^\ast  }{\sqrt{2} G_F \,M_{H_{\pm}}^2 }\,.
\end{align}
From Eqs.~\eqref{fmodel} and ~\eqref{eq:texture}, we note that there are no FCNCs in the quark sector.

 To check the compatibility of the model with direct and indirect constraints we take $g_S$ to be real, thereby maintaining CP conservation. This leaves us with only two best fit points: $g_S = 0.171$ and $g_S = -1.6$. Using $g_S = 0.171$, we find $|\lambda_{q}\lambda_{\tau \tau}^\ast| \approx 0.11$ for $M_{H_{\pm}}=200$~GeV which is consistent with direct constraints on the mass of the charged Higgs \cite{Krab:2022lih}. Alternatively, for $g_S = -1.6$, we find $|\lambda_{q}\lambda_{\tau \tau}^\ast| \approx 1.055$ for $M_{H_{\pm}} = 200$~GeV. Although it is expected that the best fit point $g_S = 0.171$ is better suited to avoid direct constraints compared to $g_S = -1.6$, we check the consistency of both points with other semileptonic decays. In the charm sector, the kinematically allowed decays are $D_s \to \tau \nu_\tau$ and $D \to \tau \nu_\tau$ but a scalar hadronic current cannot contribute to such annihilation decays as $D_s, D$ are pseudoscalars.
The NP can contribute to top decay, $t \to b \tau \nu_\tau$, but as indicated earlier, NP-SM interference is suppressed by $ {m_\tau /\sqrt{q^2}}$. Since the range of $q^2$ is $(m_t-m_b)^2 $ to $m_\tau^2$, NP effects will only be important near the endpoint of the distribution, resulting in a small correction to the  $t \to b \tau \nu_\tau$  rate. We find that $\mathcal{B}(t \to b \tau \nu_\tau)=0.110$ for $g_S = -1.6$ and $M_{H_{\pm}}=200$~GeV, which is consistent with current measurements~\cite{pdg}, while 
there is negligible impact on the rate for $g_S = 0.171$.

In a complete model, constraints from, for example, $ b \to s \gamma$ and 
$ Z \to b \bar{b}$ due to the charged Higgs loop~\cite{Degrassi:2010ne} will constrain $\lambda_q$ in Eq.~(\ref{eq:texture}). 
We calculate the corrections to the Wilson coefficient(s) $C_7^{(')}$ corresponding to the electromagnetic dipole operator(s) $\mathcal{O}_{7}^{(')} = -\frac{4G_F}{\sqrt{2}}V_{tb}V_{ts}^* \frac{e}{16\pi^2}m_b (\bar{s}_{L(R)} \sigma^{\mu \nu} b_{R(L)} F_{\mu\nu})$ that directly contribute to the $b \to s \gamma$ rate. The decay rate is directly proportional to the squares of $C_7 = C_7^{\rm eff, SM} + \delta C_7$ and $C_7^\prime = C_7^{\prime SM}+\delta C_7^\prime$. In the SM, $C_7^{\rm eff,SM} = -0.2915$ \cite{Blake:2016olu} at the scale $\mu = 4.8$ GeV which includes
next-to-next-to-leading order QCD and next-to-leading order electroweak corrections, while $C_7^{\prime SM} = \frac{m_s}{m_b}C_7 \approx 0$. The leading order corrections to the Wilson coefficients due to the charged Higgs loop are given by
\beq
\footnotesize
\delta C_7 = \delta C_7^\prime = \left(-\frac{4G_F}{\sqrt{2}}V_{tb}V_{ts}^* \frac{e}{16\pi^2}m_b \right)^{-1} V_{tb}V_{ts}^* \frac{e}{32\pi^2}m_b \frac{|\lambda_q|^2}{M_{H^\pm}^2} f(x_t)\,,
\eeq
where $f({x_t}) = \frac{-37 x_t^3 + 33 x_t^2-3 x_t + 7 +6 (3 x_t^2 + 3 x_t + 2) x_t \log(x_t)}{18 (x_t-1)^4}$ is the Inami-Lim function for the loop with $x_t = m_t^2/M_{H^\pm}^2$. Constraints on $\delta C_7^{(\prime)}$ from global fits of observables sensitive to the $b\to s \gamma$ transition~\cite{Paul:2016urs} limit $\lambda_q \lesssim 0.5$.  From Eq.~\eqref{eq:CSL}, this implies that we need $|\lambda_{\tau \tau}| \approx 0.2$ to generate $g_S = 0.171$; to obtain $g_S = -1.6$ we require a relatively large $\tau$ coupling, $|\lambda_{\tau \tau}| \approx 2$. The charged Higgs loop contribution reduces the value of $R_b  = \Gamma(Z \to b \bar b)/\Gamma(Z \to \rm hadrons)$ with respect to the SM. However, for $\lambda_q = 0.5$, the correction to the $Zb \bar b$ vertex is negligible. We also calculate the box diagram contributions (with one or two charged Higgs in the loop) to $B$-mixing. We find that for $M_{H^\pm} = 200$ GeV and $\lambda_q = 0.5$, the contribution is more than an order of magnitude smaller than the SM expectation.
Moreover, $\tau \bar{\tau}$ production  will constrain the neutral Higgs masses~\cite{Faroughy:2016osc}. It has been pointed out in Ref.~\cite{Faroughy:2016osc} that a 2HDM explanation of $R_{D^{(*)}}$ cannot be reconciled with the $pp \to \tau^+ \tau^-$ search at LHC. Although the 2HDM models discussed are different from our ansatz, we find that $g_S = 0.171$ is consistent with the recasted LHC limits for neutral scalars heavier than 250~GeV. The production of the  $ \tau \nu_\tau$ final  state through a charged Higgs contribution is expected to be small for the same reason that the correction to $\mathcal{B}(t \to b \tau \nu_\tau)$ is small. 

{\bf{Summary.}}
\label{sec:conclusions}
LHCb measurements of $\RD$, $\RDstar$ and $\Rlc$ have shown hints of lepton universality violation for years. Recent measurements, however, bring the value of $\RDstar$ in accord with the SM while the central of $\RD$ remains quite different from the SM expectation. We explored the possibility of new physics that affects $\RD$ and $\Rlc$ but not $\RDstar$ or $R_{J/\psi}$. In an effective theory framework this scenario picks out a new scalar operator with Wilson coefficient $g_S$. 
We fit $g_S$ to the latest LHCb data and made predictions for several observables that can be used to test the scenario. Although a scalar interaction improves the overall fit relative to the SM, the improvement is entirely governed by ${\RDMu}$ and ${\RlcMu}$ remains unchanged from the SM expectation. Future measurements of $\Rlc$, the tau polarization and forward-backward asymmetries, and $\langle D_4^\Lambda \rangle$ can be used to test the new interaction. Finally, we presented an ultraviolet complete two Higgs doublet model, and ensured that it satisfies constraints from charm and top decays, $b\to s \gamma$, $B$-mixing and the $Z\to b\bar b$ rate for real $g_S = 0.171$. 

\vspace{0.1in}
\noindent
{\bf Acknowledgments}. 
We thank Teppei Kitahara, Anirban Kundu, Stefan Meinel,  and 
Iguro Syuhei for useful discussions. D.M. thanks the Tata Institute of Fundamental Research
for its hospitality during the completion of this work.
A.D. is supported in part by the U.S. National Science Foundation under Grant No.~PHY-1915142. D.M. is supported in part by the U.S. Department of Energy under Grant No.~de-sc0010504.

\bibliography{ref}

\begin{thebibliography}{47}%
\makeatletter
\providecommand \@ifxundefined [1]{%
 \@ifx{#1\undefined}
}%
\providecommand \@ifnum [1]{%
 \ifnum #1\expandafter \@firstoftwo
 \else \expandafter \@secondoftwo
 \fi
}%
\providecommand \@ifx [1]{%
 \ifx #1\expandafter \@firstoftwo
 \else \expandafter \@secondoftwo
 \fi
}%
\providecommand \natexlab [1]{#1}%
\providecommand \enquote  [1]{``#1''}%
\providecommand \bibnamefont  [1]{#1}%
\providecommand \bibfnamefont [1]{#1}%
\providecommand \citenamefont [1]{#1}%
\providecommand \href@noop [0]{\@secondoftwo}%
\providecommand \href [0]{\begingroup \@sanitize@url \@href}%
\providecommand \@href[1]{\@@startlink{#1}\@@href}%
\providecommand \@@href[1]{\endgroup#1\@@endlink}%
\providecommand \@sanitize@url [0]{\catcode `\\12\catcode `\$12\catcode
  `\&12\catcode `\#12\catcode `\^12\catcode `\_12\catcode `\%12\relax}%
\providecommand \@@startlink[1]{}%
\providecommand \@@endlink[0]{}%
\providecommand \url  [0]{\begingroup\@sanitize@url \@url }%
\providecommand \@url [1]{\endgroup\@href {#1}{\urlprefix }}%
\providecommand \urlprefix  [0]{URL }%
\providecommand \Eprint [0]{\href }%
\providecommand \doibase [0]{http://dx.doi.org/}%
\providecommand \selectlanguage [0]{\@gobble}%
\providecommand \bibinfo  [0]{\@secondoftwo}%
\providecommand \bibfield  [0]{\@secondoftwo}%
\providecommand \translation [1]{[#1]}%
\providecommand \BibitemOpen [0]{}%
\providecommand \bibitemStop [0]{}%
\providecommand \bibitemNoStop [0]{.\EOS\space}%
\providecommand \EOS [0]{\spacefactor3000\relax}%
\providecommand \BibitemShut  [1]{\csname bibitem#1\endcsname}%
\let\auto@bib@innerbib\@empty
\bibitem [{\citenamefont {Lees}\ \emph {et~al.}(2012)\citenamefont {Lees} \emph
  {et~al.}}]{BaBar:2012obs}%
  \BibitemOpen
  \bibfield  {author} {\bibinfo {author} {\bibfnamefont {J.~P.}\ \bibnamefont
  {Lees}} \emph {et~al.} (\bibinfo {collaboration} {BaBar}),\ }\href {\doibase
  10.1103/PhysRevLett.109.101802} {\bibfield  {journal} {\bibinfo  {journal}
  {Phys. Rev. Lett.}\ }\textbf {\bibinfo {volume} {109}},\ \bibinfo {pages}
  {101802} (\bibinfo {year} {2012})},\ \Eprint {http://arxiv.org/abs/1205.5442}
  {arXiv:1205.5442 [hep-ex]} \BibitemShut {NoStop}%
\bibitem [{\citenamefont {Lees}\ \emph {et~al.}(2013)\citenamefont {Lees} \emph
  {et~al.}}]{BaBar:2013mob}%
  \BibitemOpen
  \bibfield  {author} {\bibinfo {author} {\bibfnamefont {J.~P.}\ \bibnamefont
  {Lees}} \emph {et~al.} (\bibinfo {collaboration} {BaBar}),\ }\href {\doibase
  10.1103/PhysRevD.88.072012} {\bibfield  {journal} {\bibinfo  {journal} {Phys.
  Rev. D}\ }\textbf {\bibinfo {volume} {88}},\ \bibinfo {pages} {072012}
  (\bibinfo {year} {2013})},\ \Eprint {http://arxiv.org/abs/1303.0571}
  {arXiv:1303.0571 [hep-ex]} \BibitemShut {NoStop}%
\bibitem [{\citenamefont {Huschle}\ \emph {et~al.}(2015)\citenamefont {Huschle}
  \emph {et~al.}}]{Belle:2015qfa}%
  \BibitemOpen
  \bibfield  {author} {\bibinfo {author} {\bibfnamefont {M.}~\bibnamefont
  {Huschle}} \emph {et~al.} (\bibinfo {collaboration} {Belle}),\ }\href
  {\doibase 10.1103/PhysRevD.92.072014} {\bibfield  {journal} {\bibinfo
  {journal} {Phys. Rev. D}\ }\textbf {\bibinfo {volume} {92}},\ \bibinfo
  {pages} {072014} (\bibinfo {year} {2015})},\ \Eprint
  {http://arxiv.org/abs/1507.03233} {arXiv:1507.03233 [hep-ex]} \BibitemShut
  {NoStop}%
\bibitem [{\citenamefont {Sato}\ \emph {et~al.}(2016)\citenamefont {Sato} \emph
  {et~al.}}]{Belle:2016ure}%
  \BibitemOpen
  \bibfield  {author} {\bibinfo {author} {\bibfnamefont {Y.}~\bibnamefont
  {Sato}} \emph {et~al.} (\bibinfo {collaboration} {Belle}),\ }\href {\doibase
  10.1103/PhysRevD.94.072007} {\bibfield  {journal} {\bibinfo  {journal} {Phys.
  Rev. D}\ }\textbf {\bibinfo {volume} {94}},\ \bibinfo {pages} {072007}
  (\bibinfo {year} {2016})},\ \Eprint {http://arxiv.org/abs/1607.07923}
  {arXiv:1607.07923 [hep-ex]} \BibitemShut {NoStop}%
\bibitem [{\citenamefont {Hirose}\ \emph {et~al.}(2017)\citenamefont {Hirose}
  \emph {et~al.}}]{Belle:2016dyj}%
  \BibitemOpen
  \bibfield  {author} {\bibinfo {author} {\bibfnamefont {S.}~\bibnamefont
  {Hirose}} \emph {et~al.} (\bibinfo {collaboration} {Belle}),\ }\href
  {\doibase 10.1103/PhysRevLett.118.211801} {\bibfield  {journal} {\bibinfo
  {journal} {Phys. Rev. Lett.}\ }\textbf {\bibinfo {volume} {118}},\ \bibinfo
  {pages} {211801} (\bibinfo {year} {2017})},\ \Eprint
  {http://arxiv.org/abs/1612.00529} {arXiv:1612.00529 [hep-ex]} \BibitemShut
  {NoStop}%
\bibitem [{\citenamefont {Hirose}\ \emph {et~al.}(2018)\citenamefont {Hirose}
  \emph {et~al.}}]{Belle:2017ilt}%
  \BibitemOpen
  \bibfield  {author} {\bibinfo {author} {\bibfnamefont {S.}~\bibnamefont
  {Hirose}} \emph {et~al.} (\bibinfo {collaboration} {Belle}),\ }\href
  {\doibase 10.1103/PhysRevD.97.012004} {\bibfield  {journal} {\bibinfo
  {journal} {Phys. Rev. D}\ }\textbf {\bibinfo {volume} {97}},\ \bibinfo
  {pages} {012004} (\bibinfo {year} {2018})},\ \Eprint
  {http://arxiv.org/abs/1709.00129} {arXiv:1709.00129 [hep-ex]} \BibitemShut
  {NoStop}%
\bibitem [{\citenamefont {Caria}\ \emph {et~al.}(2020)\citenamefont {Caria}
  \emph {et~al.}}]{Belle:2019rba}%
  \BibitemOpen
  \bibfield  {author} {\bibinfo {author} {\bibfnamefont {G.}~\bibnamefont
  {Caria}} \emph {et~al.} (\bibinfo {collaboration} {Belle}),\ }\href {\doibase
  10.1103/PhysRevLett.124.161803} {\bibfield  {journal} {\bibinfo  {journal}
  {Phys. Rev. Lett.}\ }\textbf {\bibinfo {volume} {124}},\ \bibinfo {pages}
  {161803} (\bibinfo {year} {2020})},\ \Eprint
  {http://arxiv.org/abs/1910.05864} {arXiv:1910.05864 [hep-ex]} \BibitemShut
  {NoStop}%
\bibitem [{\citenamefont {Aaij}\ \emph
  {et~al.}(2023{\natexlab{a}})\citenamefont {Aaij} \emph
  {et~al.}}]{LHCb:2023zxo}%
  \BibitemOpen
  \bibfield  {author} {\bibinfo {author} {\bibfnamefont {R.}~\bibnamefont
  {Aaij}} \emph {et~al.} (\bibinfo {collaboration} {LHCb}),\ }\href@noop {} {\
  (\bibinfo {year} {2023}{\natexlab{a}})},\ \Eprint
  {http://arxiv.org/abs/2302.02886} {arXiv:2302.02886 [hep-ex]} \BibitemShut
  {NoStop}%
\bibitem [{\citenamefont {Aaij}\ \emph
  {et~al.}(2023{\natexlab{b}})\citenamefont {Aaij} \emph
  {et~al.}}]{LHCb:2023cjr}%
  \BibitemOpen
  \bibfield  {author} {\bibinfo {author} {\bibfnamefont {R.}~\bibnamefont
  {Aaij}} \emph {et~al.} (\bibinfo {collaboration} {LHCb}),\ }\href@noop {} {\
  (\bibinfo {year} {2023}{\natexlab{b}})},\ \Eprint
  {http://arxiv.org/abs/2305.01463} {arXiv:2305.01463 [hep-ex]} \BibitemShut
  {NoStop}%
\bibitem [{\citenamefont {Aaij}\ \emph {et~al.}(2018)\citenamefont {Aaij} \emph
  {et~al.}}]{LHCb:2017vlu}%
  \BibitemOpen
  \bibfield  {author} {\bibinfo {author} {\bibfnamefont {R.}~\bibnamefont
  {Aaij}} \emph {et~al.} (\bibinfo {collaboration} {LHCb}),\ }\href {\doibase
  10.1103/PhysRevLett.120.121801} {\bibfield  {journal} {\bibinfo  {journal}
  {Phys. Rev. Lett.}\ }\textbf {\bibinfo {volume} {120}},\ \bibinfo {pages}
  {121801} (\bibinfo {year} {2018})},\ \Eprint
  {http://arxiv.org/abs/1711.05623} {arXiv:1711.05623 [hep-ex]} \BibitemShut
  {NoStop}%
\bibitem [{HFL(2021)}]{HFLAV-WA-2021}%
  \BibitemOpen
  \href@noop {} {\enquote {\bibinfo {title} {Average of $r(d)$ and $r(d^*)$ for
  2021},}\ }\bibinfo {howpublished}
  {\url{https://hflav-eos.web.cern.ch/hflav-eos/semi/spring21/html/RDsDsstar/RDRDs.html}}
  (\bibinfo {year} {2021})\BibitemShut {NoStop}%
\bibitem [{\citenamefont {Detmold}\ \emph {et~al.}(2015)\citenamefont
  {Detmold}, \citenamefont {Lehner},\ and\ \citenamefont
  {Meinel}}]{Detmold:2015aaa}%
  \BibitemOpen
  \bibfield  {author} {\bibinfo {author} {\bibfnamefont {W.}~\bibnamefont
  {Detmold}}, \bibinfo {author} {\bibfnamefont {C.}~\bibnamefont {Lehner}}, \
  and\ \bibinfo {author} {\bibfnamefont {S.}~\bibnamefont {Meinel}},\ }\href
  {\doibase 10.1103/PhysRevD.92.034503} {\bibfield  {journal} {\bibinfo
  {journal} {Phys. Rev. D}\ }\textbf {\bibinfo {volume} {92}},\ \bibinfo
  {pages} {034503} (\bibinfo {year} {2015})},\ \Eprint
  {http://arxiv.org/abs/1503.01421} {arXiv:1503.01421 [hep-lat]} \BibitemShut
  {NoStop}%
\bibitem [{\citenamefont {Aaij}\ \emph {et~al.}(2022)\citenamefont {Aaij} \emph
  {et~al.}}]{LHCb:2022piu}%
  \BibitemOpen
  \bibfield  {author} {\bibinfo {author} {\bibfnamefont {R.}~\bibnamefont
  {Aaij}} \emph {et~al.} (\bibinfo {collaboration} {LHCb}),\ }\href {\doibase
  10.1103/PhysRevLett.128.191803} {\bibfield  {journal} {\bibinfo  {journal}
  {Phys. Rev. Lett.}\ }\textbf {\bibinfo {volume} {128}},\ \bibinfo {pages}
  {191803} (\bibinfo {year} {2022})},\ \Eprint
  {http://arxiv.org/abs/2201.03497} {arXiv:2201.03497 [hep-ex]} \BibitemShut
  {NoStop}%
\bibitem [{\citenamefont {Bernlochner}\ \emph {et~al.}(2023)\citenamefont
  {Bernlochner}, \citenamefont {Ligeti}, \citenamefont {Papucci},\ and\
  \citenamefont {Robinson}}]{Bernlochner:2022hyz}%
  \BibitemOpen
  \bibfield  {author} {\bibinfo {author} {\bibfnamefont {F.~U.}\ \bibnamefont
  {Bernlochner}}, \bibinfo {author} {\bibfnamefont {Z.}~\bibnamefont {Ligeti}},
  \bibinfo {author} {\bibfnamefont {M.}~\bibnamefont {Papucci}}, \ and\
  \bibinfo {author} {\bibfnamefont {D.~J.}\ \bibnamefont {Robinson}},\ }\href
  {\doibase 10.1103/PhysRevD.107.L011502} {\bibfield  {journal} {\bibinfo
  {journal} {Phys. Rev. D}\ }\textbf {\bibinfo {volume} {107}},\ \bibinfo
  {pages} {L011502} (\bibinfo {year} {2023})},\ \Eprint
  {http://arxiv.org/abs/2206.11282} {arXiv:2206.11282 [hep-ph]} \BibitemShut
  {NoStop}%
\bibitem [{\citenamefont {Bernlochner}\ \emph {et~al.}(2018)\citenamefont
  {Bernlochner}, \citenamefont {Ligeti}, \citenamefont {Robinson},\ and\
  \citenamefont {Sutcliffe}}]{Bernlochner:2018kxh}%
  \BibitemOpen
  \bibfield  {author} {\bibinfo {author} {\bibfnamefont {F.~U.}\ \bibnamefont
  {Bernlochner}}, \bibinfo {author} {\bibfnamefont {Z.}~\bibnamefont {Ligeti}},
  \bibinfo {author} {\bibfnamefont {D.~J.}\ \bibnamefont {Robinson}}, \ and\
  \bibinfo {author} {\bibfnamefont {W.~L.}\ \bibnamefont {Sutcliffe}},\ }\href
  {\doibase 10.1103/PhysRevLett.121.202001} {\bibfield  {journal} {\bibinfo
  {journal} {Phys. Rev. Lett.}\ }\textbf {\bibinfo {volume} {121}},\ \bibinfo
  {pages} {202001} (\bibinfo {year} {2018})},\ \Eprint
  {http://arxiv.org/abs/1808.09464} {arXiv:1808.09464 [hep-ph]} \BibitemShut
  {NoStop}%
\bibitem [{\citenamefont {Bernlochner}\ \emph {et~al.}(2019)\citenamefont
  {Bernlochner}, \citenamefont {Ligeti}, \citenamefont {Robinson},\ and\
  \citenamefont {Sutcliffe}}]{Bernlochner:2018bfn}%
  \BibitemOpen
  \bibfield  {author} {\bibinfo {author} {\bibfnamefont {F.~U.}\ \bibnamefont
  {Bernlochner}}, \bibinfo {author} {\bibfnamefont {Z.}~\bibnamefont {Ligeti}},
  \bibinfo {author} {\bibfnamefont {D.~J.}\ \bibnamefont {Robinson}}, \ and\
  \bibinfo {author} {\bibfnamefont {W.~L.}\ \bibnamefont {Sutcliffe}},\ }\href
  {\doibase 10.1103/PhysRevD.99.055008} {\bibfield  {journal} {\bibinfo
  {journal} {Phys. Rev. D}\ }\textbf {\bibinfo {volume} {99}},\ \bibinfo
  {pages} {055008} (\bibinfo {year} {2019})},\ \Eprint
  {http://arxiv.org/abs/1812.07593} {arXiv:1812.07593 [hep-ph]} \BibitemShut
  {NoStop}%
\bibitem [{\citenamefont {Murgui}\ \emph {et~al.}(2019)\citenamefont {Murgui},
  \citenamefont {Pe{\~ n}uelas}, \citenamefont {Jung},\ and\ \citenamefont
  {Pich}}]{Murgui:2019czp}%
  \BibitemOpen
  \bibfield  {author} {\bibinfo {author} {\bibfnamefont {C.}~\bibnamefont
  {Murgui}}, \bibinfo {author} {\bibfnamefont {A.}~\bibnamefont {Pe{\~
  n}uelas}}, \bibinfo {author} {\bibfnamefont {M.}~\bibnamefont {Jung}}, \ and\
  \bibinfo {author} {\bibfnamefont {A.}~\bibnamefont {Pich}},\ }\href {\doibase
  10.1007/JHEP09(2019)103} {\bibfield  {journal} {\bibinfo  {journal} {JHEP}\
  }\textbf {\bibinfo {volume} {09}},\ \bibinfo {pages} {103} (\bibinfo {year}
  {2019})},\ \Eprint {http://arxiv.org/abs/1904.09311} {arXiv:1904.09311
  [hep-ph]} \BibitemShut {NoStop}%
\bibitem [{\citenamefont {Iguro}\ \emph {et~al.}(2022)\citenamefont {Iguro},
  \citenamefont {Kitahara},\ and\ \citenamefont {Watanabe}}]{Iguro:2022yzr}%
  \BibitemOpen
  \bibfield  {author} {\bibinfo {author} {\bibfnamefont {S.}~\bibnamefont
  {Iguro}}, \bibinfo {author} {\bibfnamefont {T.}~\bibnamefont {Kitahara}}, \
  and\ \bibinfo {author} {\bibfnamefont {R.}~\bibnamefont {Watanabe}},\
  }\href@noop {} {\  (\bibinfo {year} {2022})},\ \Eprint
  {http://arxiv.org/abs/2210.10751} {arXiv:2210.10751 [hep-ph]} \BibitemShut
  {NoStop}%
\bibitem [{\citenamefont {Ray}\ and\ \citenamefont
  {Nandi}(2023)}]{Ray:2023xjn}%
  \BibitemOpen
  \bibfield  {author} {\bibinfo {author} {\bibfnamefont {I.}~\bibnamefont
  {Ray}}\ and\ \bibinfo {author} {\bibfnamefont {S.}~\bibnamefont {Nandi}},\
  }\href@noop {} {\  (\bibinfo {year} {2023})},\ \Eprint
  {http://arxiv.org/abs/2305.11855} {arXiv:2305.11855 [hep-ph]} \BibitemShut
  {NoStop}%
\bibitem [{\citenamefont {Dumont}\ \emph {et~al.}(2016)\citenamefont {Dumont},
  \citenamefont {Nishiwaki},\ and\ \citenamefont {Watanabe}}]{Dumont:2016xpj}%
  \BibitemOpen
  \bibfield  {author} {\bibinfo {author} {\bibfnamefont {B.}~\bibnamefont
  {Dumont}}, \bibinfo {author} {\bibfnamefont {K.}~\bibnamefont {Nishiwaki}}, \
  and\ \bibinfo {author} {\bibfnamefont {R.}~\bibnamefont {Watanabe}},\ }\href
  {\doibase 10.1103/PhysRevD.94.034001} {\bibfield  {journal} {\bibinfo
  {journal} {Phys. Rev. D}\ }\textbf {\bibinfo {volume} {94}},\ \bibinfo
  {pages} {034001} (\bibinfo {year} {2016})},\ \Eprint
  {http://arxiv.org/abs/1603.05248} {arXiv:1603.05248 [hep-ph]} \BibitemShut
  {NoStop}%
\bibitem [{\citenamefont {Chen}\ and\ \citenamefont
  {Geng}(2005)}]{Chen:2005gr}%
  \BibitemOpen
  \bibfield  {author} {\bibinfo {author} {\bibfnamefont {C.-H.}\ \bibnamefont
  {Chen}}\ and\ \bibinfo {author} {\bibfnamefont {C.-Q.}\ \bibnamefont
  {Geng}},\ }\href {\doibase 10.1103/PhysRevD.71.077501} {\bibfield  {journal}
  {\bibinfo  {journal} {Phys. Rev. D}\ }\textbf {\bibinfo {volume} {71}},\
  \bibinfo {pages} {077501} (\bibinfo {year} {2005})},\ \Eprint
  {http://arxiv.org/abs/hep-ph/0503123} {arXiv:hep-ph/0503123} \BibitemShut
  {NoStop}%
\bibitem [{\citenamefont {Bhattacharya}\ \emph {et~al.}(2012)\citenamefont
  {Bhattacharya}, \citenamefont {Cirigliano}, \citenamefont {Cohen},
  \citenamefont {Filipuzzi}, \citenamefont {Gonzalez-Alonso}, \citenamefont
  {Graesser}, \citenamefont {Gupta},\ and\ \citenamefont
  {Lin}}]{Bhattacharya:2011qm}%
  \BibitemOpen
  \bibfield  {author} {\bibinfo {author} {\bibfnamefont {T.}~\bibnamefont
  {Bhattacharya}}, \bibinfo {author} {\bibfnamefont {V.}~\bibnamefont
  {Cirigliano}}, \bibinfo {author} {\bibfnamefont {S.~D.}\ \bibnamefont
  {Cohen}}, \bibinfo {author} {\bibfnamefont {A.}~\bibnamefont {Filipuzzi}},
  \bibinfo {author} {\bibfnamefont {M.}~\bibnamefont {Gonzalez-Alonso}},
  \bibinfo {author} {\bibfnamefont {M.~L.}\ \bibnamefont {Graesser}}, \bibinfo
  {author} {\bibfnamefont {R.}~\bibnamefont {Gupta}}, \ and\ \bibinfo {author}
  {\bibfnamefont {H.-W.}\ \bibnamefont {Lin}},\ }\href {\doibase
  10.1103/PhysRevD.85.054512} {\bibfield  {journal} {\bibinfo  {journal} {Phys.
  Rev. D}\ }\textbf {\bibinfo {volume} {85}},\ \bibinfo {pages} {054512}
  (\bibinfo {year} {2012})},\ \Eprint {http://arxiv.org/abs/1110.6448}
  {arXiv:1110.6448 [hep-ph]} \BibitemShut {NoStop}%
\bibitem [{\citenamefont {Feruglio}\ \emph
  {et~al.}(2017{\natexlab{a}})\citenamefont {Feruglio}, \citenamefont
  {Paradisi},\ and\ \citenamefont {Pattori}}]{Feruglio:2016gvd}%
  \BibitemOpen
  \bibfield  {author} {\bibinfo {author} {\bibfnamefont {F.}~\bibnamefont
  {Feruglio}}, \bibinfo {author} {\bibfnamefont {P.}~\bibnamefont {Paradisi}},
  \ and\ \bibinfo {author} {\bibfnamefont {A.}~\bibnamefont {Pattori}},\ }\href
  {\doibase 10.1103/PhysRevLett.118.011801} {\bibfield  {journal} {\bibinfo
  {journal} {Phys. Rev. Lett.}\ }\textbf {\bibinfo {volume} {118}},\ \bibinfo
  {pages} {011801} (\bibinfo {year} {2017}{\natexlab{a}})},\ \Eprint
  {http://arxiv.org/abs/1606.00524} {arXiv:1606.00524 [hep-ph]} \BibitemShut
  {NoStop}%
\bibitem [{\citenamefont {Feruglio}\ \emph
  {et~al.}(2017{\natexlab{b}})\citenamefont {Feruglio}, \citenamefont
  {Paradisi},\ and\ \citenamefont {Pattori}}]{Feruglio:2017rjo}%
  \BibitemOpen
  \bibfield  {author} {\bibinfo {author} {\bibfnamefont {F.}~\bibnamefont
  {Feruglio}}, \bibinfo {author} {\bibfnamefont {P.}~\bibnamefont {Paradisi}},
  \ and\ \bibinfo {author} {\bibfnamefont {A.}~\bibnamefont {Pattori}},\ }\href
  {\doibase 10.1007/JHEP09(2017)061} {\bibfield  {journal} {\bibinfo  {journal}
  {JHEP}\ }\textbf {\bibinfo {volume} {09}},\ \bibinfo {pages} {061} (\bibinfo
  {year} {2017}{\natexlab{b}})},\ \Eprint {http://arxiv.org/abs/1705.00929}
  {arXiv:1705.00929 [hep-ph]} \BibitemShut {NoStop}%
\bibitem [{\citenamefont {Be\v{c}irevi\'c}\ and\ \citenamefont
  {Jaffredo}(2022)}]{Becirevic:2022bev}%
  \BibitemOpen
  \bibfield  {author} {\bibinfo {author} {\bibfnamefont {D.}~\bibnamefont
  {Be\v{c}irevi\'c}}\ and\ \bibinfo {author} {\bibfnamefont {F.}~\bibnamefont
  {Jaffredo}},\ }\href@noop {} {\  (\bibinfo {year} {2022})},\ \Eprint
  {http://arxiv.org/abs/2209.13409} {arXiv:2209.13409 [hep-ph]} \BibitemShut
  {NoStop}%
\bibitem [{\citenamefont {Na}\ \emph {et~al.}(2015)\citenamefont {Na},
  \citenamefont {Bouchard}, \citenamefont {Lepage}, \citenamefont {Monahan},\
  and\ \citenamefont {Shigemitsu}}]{Na:2015kha}%
  \BibitemOpen
  \bibfield  {author} {\bibinfo {author} {\bibfnamefont {H.}~\bibnamefont
  {Na}}, \bibinfo {author} {\bibfnamefont {C.~M.}\ \bibnamefont {Bouchard}},
  \bibinfo {author} {\bibfnamefont {G.~P.}\ \bibnamefont {Lepage}}, \bibinfo
  {author} {\bibfnamefont {C.}~\bibnamefont {Monahan}}, \ and\ \bibinfo
  {author} {\bibfnamefont {J.}~\bibnamefont {Shigemitsu}} (\bibinfo
  {collaboration} {HPQCD}),\ }\href {\doibase 10.1103/PhysRevD.93.119906}
  {\bibfield  {journal} {\bibinfo  {journal} {Phys. Rev. D}\ }\textbf {\bibinfo
  {volume} {92}},\ \bibinfo {pages} {054510} (\bibinfo {year} {2015})},\
  \bibinfo {note} {[Erratum: Phys.Rev.D 93, 119906 (2016)]},\ \Eprint
  {http://arxiv.org/abs/1505.03925} {arXiv:1505.03925 [hep-lat]} \BibitemShut
  {NoStop}%
\bibitem [{\citenamefont {Bailey}\ \emph {et~al.}(2015)\citenamefont {Bailey}
  \emph {et~al.}}]{MILC:2015uhg}%
  \BibitemOpen
  \bibfield  {author} {\bibinfo {author} {\bibfnamefont {J.~A.}\ \bibnamefont
  {Bailey}} \emph {et~al.} (\bibinfo {collaboration} {MILC}),\ }\href {\doibase
  10.1103/PhysRevD.92.034506} {\bibfield  {journal} {\bibinfo  {journal} {Phys.
  Rev. D}\ }\textbf {\bibinfo {volume} {92}},\ \bibinfo {pages} {034506}
  (\bibinfo {year} {2015})},\ \Eprint {http://arxiv.org/abs/1503.07237}
  {arXiv:1503.07237 [hep-lat]} \BibitemShut {NoStop}%
\bibitem [{\citenamefont {Boyd}\ \emph {et~al.}(1997)\citenamefont {Boyd},
  \citenamefont {Grinstein},\ and\ \citenamefont {Lebed}}]{PhysRevD.56.6895}%
  \BibitemOpen
  \bibfield  {author} {\bibinfo {author} {\bibfnamefont {C.~G.}\ \bibnamefont
  {Boyd}}, \bibinfo {author} {\bibfnamefont {B.}~\bibnamefont {Grinstein}}, \
  and\ \bibinfo {author} {\bibfnamefont {R.~F.}\ \bibnamefont {Lebed}},\ }\href
  {\doibase 10.1103/PhysRevD.56.6895} {\bibfield  {journal} {\bibinfo
  {journal} {Phys. Rev. D}\ }\textbf {\bibinfo {volume} {56}},\ \bibinfo
  {pages} {6895} (\bibinfo {year} {1997})}\BibitemShut {NoStop}%
\bibitem [{\citenamefont {Datta}\ \emph {et~al.}(2017)\citenamefont {Datta},
  \citenamefont {Kamali}, \citenamefont {Meinel},\ and\ \citenamefont
  {Rashed}}]{Datta:2017aue}%
  \BibitemOpen
  \bibfield  {author} {\bibinfo {author} {\bibfnamefont {A.}~\bibnamefont
  {Datta}}, \bibinfo {author} {\bibfnamefont {S.}~\bibnamefont {Kamali}},
  \bibinfo {author} {\bibfnamefont {S.}~\bibnamefont {Meinel}}, \ and\ \bibinfo
  {author} {\bibfnamefont {A.}~\bibnamefont {Rashed}},\ }\href {\doibase
  10.1007/JHEP08(2017)131} {\bibfield  {journal} {\bibinfo  {journal} {JHEP}\
  }\textbf {\bibinfo {volume} {08}},\ \bibinfo {pages} {131} (\bibinfo {year}
  {2017})},\ \Eprint {http://arxiv.org/abs/1702.02243} {arXiv:1702.02243
  [hep-ph]} \BibitemShut {NoStop}%
\bibitem [{\citenamefont {Bourrely}\ \emph {et~al.}(2009)\citenamefont
  {Bourrely}, \citenamefont {Caprini},\ and\ \citenamefont
  {Lellouch}}]{Bourrely:2008za}%
  \BibitemOpen
  \bibfield  {author} {\bibinfo {author} {\bibfnamefont {C.}~\bibnamefont
  {Bourrely}}, \bibinfo {author} {\bibfnamefont {I.}~\bibnamefont {Caprini}}, \
  and\ \bibinfo {author} {\bibfnamefont {L.}~\bibnamefont {Lellouch}},\ }\href
  {\doibase 10.1103/PhysRevD.82.099902} {\bibfield  {journal} {\bibinfo
  {journal} {Phys. Rev. D}\ }\textbf {\bibinfo {volume} {79}},\ \bibinfo
  {pages} {013008} (\bibinfo {year} {2009})},\ \bibinfo {note} {[Erratum:
  Phys.Rev.D 82, 099902 (2010)]},\ \Eprint {http://arxiv.org/abs/0807.2722}
  {arXiv:0807.2722 [hep-ph]} \BibitemShut {NoStop}%
\bibitem [{\citenamefont {Fedele}\ \emph {et~al.}(2022)\citenamefont {Fedele},
  \citenamefont {Blanke}, \citenamefont {Crivellin}, \citenamefont {Iguro},
  \citenamefont {Kitahara}, \citenamefont {Nierste},\ and\ \citenamefont
  {Watanabe}}]{Fedele:2022iib}%
  \BibitemOpen
  \bibfield  {author} {\bibinfo {author} {\bibfnamefont {M.}~\bibnamefont
  {Fedele}}, \bibinfo {author} {\bibfnamefont {M.}~\bibnamefont {Blanke}},
  \bibinfo {author} {\bibfnamefont {A.}~\bibnamefont {Crivellin}}, \bibinfo
  {author} {\bibfnamefont {S.}~\bibnamefont {Iguro}}, \bibinfo {author}
  {\bibfnamefont {T.}~\bibnamefont {Kitahara}}, \bibinfo {author}
  {\bibfnamefont {U.}~\bibnamefont {Nierste}}, \ and\ \bibinfo {author}
  {\bibfnamefont {R.}~\bibnamefont {Watanabe}},\ }\href@noop {} {\  (\bibinfo
  {year} {2022})},\ \Eprint {http://arxiv.org/abs/2211.14172} {arXiv:2211.14172
  [hep-ph]} \BibitemShut {NoStop}%
\bibitem [{\citenamefont {Lee}(1973)}]{Lee:1973iz}%
  \BibitemOpen
  \bibfield  {author} {\bibinfo {author} {\bibfnamefont {T.~D.}\ \bibnamefont
  {Lee}},\ }\href {\doibase 10.1103/PhysRevD.8.1226} {\bibfield  {journal}
  {\bibinfo  {journal} {Phys. Rev. D}\ }\textbf {\bibinfo {volume} {8}},\
  \bibinfo {pages} {1226} (\bibinfo {year} {1973})}\BibitemShut {NoStop}%
\bibitem [{\citenamefont {Branco}\ \emph {et~al.}(2012)\citenamefont {Branco},
  \citenamefont {Ferreira}, \citenamefont {Lavoura}, \citenamefont {Rebelo},
  \citenamefont {Sher},\ and\ \citenamefont {Silva}}]{Branco:2011iw}%
  \BibitemOpen
  \bibfield  {author} {\bibinfo {author} {\bibfnamefont {G.~C.}\ \bibnamefont
  {Branco}}, \bibinfo {author} {\bibfnamefont {P.~M.}\ \bibnamefont
  {Ferreira}}, \bibinfo {author} {\bibfnamefont {L.}~\bibnamefont {Lavoura}},
  \bibinfo {author} {\bibfnamefont {M.~N.}\ \bibnamefont {Rebelo}}, \bibinfo
  {author} {\bibfnamefont {M.}~\bibnamefont {Sher}}, \ and\ \bibinfo {author}
  {\bibfnamefont {J.~P.}\ \bibnamefont {Silva}},\ }\href {\doibase
  10.1016/j.physrep.2012.02.002} {\bibfield  {journal} {\bibinfo  {journal}
  {Phys. Rept.}\ }\textbf {\bibinfo {volume} {516}},\ \bibinfo {pages} {1}
  (\bibinfo {year} {2012})},\ \Eprint {http://arxiv.org/abs/1106.0034}
  {arXiv:1106.0034 [hep-ph]} \BibitemShut {NoStop}%
\bibitem [{\citenamefont {Davidson}\ and\ \citenamefont
  {Haber}(2005)}]{Davidson:2005cw}%
  \BibitemOpen
  \bibfield  {author} {\bibinfo {author} {\bibfnamefont {S.}~\bibnamefont
  {Davidson}}\ and\ \bibinfo {author} {\bibfnamefont {H.~E.}\ \bibnamefont
  {Haber}},\ }\href {\doibase 10.1103/PhysRevD.72.099902} {\bibfield  {journal}
  {\bibinfo  {journal} {Phys. Rev. D}\ }\textbf {\bibinfo {volume} {72}},\
  \bibinfo {pages} {035004} (\bibinfo {year} {2005})},\ \bibinfo {note}
  {[Erratum: Phys.Rev.D 72, 099902 (2005)]},\ \Eprint
  {http://arxiv.org/abs/hep-ph/0504050} {arXiv:hep-ph/0504050} \BibitemShut
  {NoStop}%
\bibitem [{\citenamefont {Pakvasa}\ and\ \citenamefont
  {Sugawara}(1978)}]{Pakvasa:1977in}%
  \BibitemOpen
  \bibfield  {author} {\bibinfo {author} {\bibfnamefont {S.}~\bibnamefont
  {Pakvasa}}\ and\ \bibinfo {author} {\bibfnamefont {H.}~\bibnamefont
  {Sugawara}},\ }\href {\doibase 10.1016/0370-2693(78)90172-7} {\bibfield
  {journal} {\bibinfo  {journal} {Phys. Lett. B}\ }\textbf {\bibinfo {volume}
  {73}},\ \bibinfo {pages} {61} (\bibinfo {year} {1978})}\BibitemShut {NoStop}%
\bibitem [{\citenamefont {Atwood}\ \emph {et~al.}(1997)\citenamefont {Atwood},
  \citenamefont {Reina},\ and\ \citenamefont {Soni}}]{Atwood:1996vj}%
  \BibitemOpen
  \bibfield  {author} {\bibinfo {author} {\bibfnamefont {D.}~\bibnamefont
  {Atwood}}, \bibinfo {author} {\bibfnamefont {L.}~\bibnamefont {Reina}}, \
  and\ \bibinfo {author} {\bibfnamefont {A.}~\bibnamefont {Soni}},\ }\href
  {\doibase 10.1103/PhysRevD.55.3156} {\bibfield  {journal} {\bibinfo
  {journal} {Phys. Rev. D}\ }\textbf {\bibinfo {volume} {55}},\ \bibinfo
  {pages} {3156} (\bibinfo {year} {1997})},\ \Eprint
  {http://arxiv.org/abs/hep-ph/9609279} {arXiv:hep-ph/9609279} \BibitemShut
  {NoStop}%
\bibitem [{\citenamefont {Alonso}\ \emph {et~al.}(2017)\citenamefont {Alonso},
  \citenamefont {Grinstein},\ and\ \citenamefont
  {Martin~Camalich}}]{Alonso:2016oyd}%
  \BibitemOpen
  \bibfield  {author} {\bibinfo {author} {\bibfnamefont {R.}~\bibnamefont
  {Alonso}}, \bibinfo {author} {\bibfnamefont {B.}~\bibnamefont {Grinstein}}, \
  and\ \bibinfo {author} {\bibfnamefont {J.}~\bibnamefont {Martin~Camalich}},\
  }\href {\doibase 10.1103/PhysRevLett.118.081802} {\bibfield  {journal}
  {\bibinfo  {journal} {Phys. Rev. Lett.}\ }\textbf {\bibinfo {volume} {118}},\
  \bibinfo {pages} {081802} (\bibinfo {year} {2017})},\ \Eprint
  {http://arxiv.org/abs/1611.06676} {arXiv:1611.06676 [hep-ph]} \BibitemShut
  {NoStop}%
\bibitem [{\citenamefont {Fajfer}\ \emph {et~al.}(2012)\citenamefont {Fajfer},
  \citenamefont {Kamenik}, \citenamefont {Nisandzic},\ and\ \citenamefont
  {Zupan}}]{Fajfer:2012jt}%
  \BibitemOpen
  \bibfield  {author} {\bibinfo {author} {\bibfnamefont {S.}~\bibnamefont
  {Fajfer}}, \bibinfo {author} {\bibfnamefont {J.~F.}\ \bibnamefont {Kamenik}},
  \bibinfo {author} {\bibfnamefont {I.}~\bibnamefont {Nisandzic}}, \ and\
  \bibinfo {author} {\bibfnamefont {J.}~\bibnamefont {Zupan}},\ }\href
  {\doibase 10.1103/PhysRevLett.109.161801} {\bibfield  {journal} {\bibinfo
  {journal} {Phys. Rev. Lett.}\ }\textbf {\bibinfo {volume} {109}},\ \bibinfo
  {pages} {161801} (\bibinfo {year} {2012})},\ \Eprint
  {http://arxiv.org/abs/1206.1872} {arXiv:1206.1872 [hep-ph]} \BibitemShut
  {NoStop}%
\bibitem [{\citenamefont {Crivellin}\ \emph {et~al.}(2012)\citenamefont
  {Crivellin}, \citenamefont {Greub},\ and\ \citenamefont
  {Kokulu}}]{Crivellin:2012ye}%
  \BibitemOpen
  \bibfield  {author} {\bibinfo {author} {\bibfnamefont {A.}~\bibnamefont
  {Crivellin}}, \bibinfo {author} {\bibfnamefont {C.}~\bibnamefont {Greub}}, \
  and\ \bibinfo {author} {\bibfnamefont {A.}~\bibnamefont {Kokulu}},\ }\href
  {\doibase 10.1103/PhysRevD.86.054014} {\bibfield  {journal} {\bibinfo
  {journal} {Phys. Rev. D}\ }\textbf {\bibinfo {volume} {86}},\ \bibinfo
  {pages} {054014} (\bibinfo {year} {2012})},\ \Eprint
  {http://arxiv.org/abs/1206.2634} {arXiv:1206.2634 [hep-ph]} \BibitemShut
  {NoStop}%
\bibitem [{\citenamefont {Celis}\ \emph {et~al.}(2013)\citenamefont {Celis},
  \citenamefont {Jung}, \citenamefont {Li},\ and\ \citenamefont
  {Pich}}]{Celis:2012dk}%
  \BibitemOpen
  \bibfield  {author} {\bibinfo {author} {\bibfnamefont {A.}~\bibnamefont
  {Celis}}, \bibinfo {author} {\bibfnamefont {M.}~\bibnamefont {Jung}},
  \bibinfo {author} {\bibfnamefont {X.-Q.}\ \bibnamefont {Li}}, \ and\ \bibinfo
  {author} {\bibfnamefont {A.}~\bibnamefont {Pich}},\ }\href {\doibase
  10.1007/JHEP01(2013)054} {\bibfield  {journal} {\bibinfo  {journal} {JHEP}\
  }\textbf {\bibinfo {volume} {01}},\ \bibinfo {pages} {054} (\bibinfo {year}
  {2013})},\ \Eprint {http://arxiv.org/abs/1210.8443} {arXiv:1210.8443
  [hep-ph]} \BibitemShut {NoStop}%
\bibitem [{\citenamefont {Pich}\ and\ \citenamefont
  {Tuzon}(2009)}]{Pich:2009sp}%
  \BibitemOpen
  \bibfield  {author} {\bibinfo {author} {\bibfnamefont {A.}~\bibnamefont
  {Pich}}\ and\ \bibinfo {author} {\bibfnamefont {P.}~\bibnamefont {Tuzon}},\
  }\href {\doibase 10.1103/PhysRevD.80.091702} {\bibfield  {journal} {\bibinfo
  {journal} {Phys. Rev. D}\ }\textbf {\bibinfo {volume} {80}},\ \bibinfo
  {pages} {091702} (\bibinfo {year} {2009})},\ \Eprint
  {http://arxiv.org/abs/0908.1554} {arXiv:0908.1554 [hep-ph]} \BibitemShut
  {NoStop}%
\bibitem [{\citenamefont {Krab}\ \emph {et~al.}(2023)\citenamefont {Krab},
  \citenamefont {Ouchemhou}, \citenamefont {Arhrib}, \citenamefont {Benbrik},
  \citenamefont {Manaut},\ and\ \citenamefont {Yan}}]{Krab:2022lih}%
  \BibitemOpen
  \bibfield  {author} {\bibinfo {author} {\bibfnamefont {M.}~\bibnamefont
  {Krab}}, \bibinfo {author} {\bibfnamefont {M.}~\bibnamefont {Ouchemhou}},
  \bibinfo {author} {\bibfnamefont {A.}~\bibnamefont {Arhrib}}, \bibinfo
  {author} {\bibfnamefont {R.}~\bibnamefont {Benbrik}}, \bibinfo {author}
  {\bibfnamefont {B.}~\bibnamefont {Manaut}}, \ and\ \bibinfo {author}
  {\bibfnamefont {Q.-S.}\ \bibnamefont {Yan}},\ }\href {\doibase
  10.1016/j.physletb.2023.137705} {\bibfield  {journal} {\bibinfo  {journal}
  {Phys. Lett. B}\ }\textbf {\bibinfo {volume} {839}},\ \bibinfo {pages}
  {137705} (\bibinfo {year} {2023})},\ \Eprint
  {http://arxiv.org/abs/2210.09416} {arXiv:2210.09416 [hep-ph]} \BibitemShut
  {NoStop}%
\bibitem [{\citenamefont {Zyla}\ \emph {et~al.}(2020)\citenamefont {Zyla} \emph
  {et~al.}}]{pdg}%
  \BibitemOpen
  \bibfield  {author} {\bibinfo {author} {\bibfnamefont {P.}~\bibnamefont
  {Zyla}} \emph {et~al.} (\bibinfo {collaboration} {Particle Data Group}),\
  }\href {\doibase 10.1093/ptep/ptaa104} {\bibfield  {journal} {\bibinfo
  {journal} {PTEP}\ }\textbf {\bibinfo {volume} {2020}},\ \bibinfo {pages}
  {083C01} (\bibinfo {year} {2020})}\BibitemShut {NoStop}%
\bibitem [{\citenamefont {Degrassi}\ and\ \citenamefont
  {Slavich}(2010)}]{Degrassi:2010ne}%
  \BibitemOpen
  \bibfield  {author} {\bibinfo {author} {\bibfnamefont {G.}~\bibnamefont
  {Degrassi}}\ and\ \bibinfo {author} {\bibfnamefont {P.}~\bibnamefont
  {Slavich}},\ }\href {\doibase 10.1103/PhysRevD.81.075001} {\bibfield
  {journal} {\bibinfo  {journal} {Phys. Rev. D}\ }\textbf {\bibinfo {volume}
  {81}},\ \bibinfo {pages} {075001} (\bibinfo {year} {2010})},\ \Eprint
  {http://arxiv.org/abs/1002.1071} {arXiv:1002.1071 [hep-ph]} \BibitemShut
  {NoStop}%
\bibitem [{\citenamefont {Blake}\ \emph {et~al.}(2017)\citenamefont {Blake},
  \citenamefont {Lanfranchi},\ and\ \citenamefont {Straub}}]{Blake:2016olu}%
  \BibitemOpen
  \bibfield  {author} {\bibinfo {author} {\bibfnamefont {T.}~\bibnamefont
  {Blake}}, \bibinfo {author} {\bibfnamefont {G.}~\bibnamefont {Lanfranchi}}, \
  and\ \bibinfo {author} {\bibfnamefont {D.~M.}\ \bibnamefont {Straub}},\
  }\href {\doibase 10.1016/j.ppnp.2016.10.001} {\bibfield  {journal} {\bibinfo
  {journal} {Prog. Part. Nucl. Phys.}\ }\textbf {\bibinfo {volume} {92}},\
  \bibinfo {pages} {50} (\bibinfo {year} {2017})},\ \Eprint
  {http://arxiv.org/abs/1606.00916} {arXiv:1606.00916 [hep-ph]} \BibitemShut
  {NoStop}%
\bibitem [{\citenamefont {Paul}\ and\ \citenamefont
  {Straub}(2017)}]{Paul:2016urs}%
  \BibitemOpen
  \bibfield  {author} {\bibinfo {author} {\bibfnamefont {A.}~\bibnamefont
  {Paul}}\ and\ \bibinfo {author} {\bibfnamefont {D.~M.}\ \bibnamefont
  {Straub}},\ }\href {\doibase 10.1007/JHEP04(2017)027} {\bibfield  {journal}
  {\bibinfo  {journal} {JHEP}\ }\textbf {\bibinfo {volume} {04}},\ \bibinfo
  {pages} {027} (\bibinfo {year} {2017})},\ \Eprint
  {http://arxiv.org/abs/1608.02556} {arXiv:1608.02556 [hep-ph]} \BibitemShut
  {NoStop}%
\bibitem [{\citenamefont {Faroughy}\ \emph {et~al.}(2017)\citenamefont
  {Faroughy}, \citenamefont {Greljo},\ and\ \citenamefont
  {Kamenik}}]{Faroughy:2016osc}%
  \BibitemOpen
  \bibfield  {author} {\bibinfo {author} {\bibfnamefont {D.~A.}\ \bibnamefont
  {Faroughy}}, \bibinfo {author} {\bibfnamefont {A.}~\bibnamefont {Greljo}}, \
  and\ \bibinfo {author} {\bibfnamefont {J.~F.}\ \bibnamefont {Kamenik}},\
  }\href {\doibase 10.1016/j.physletb.2016.11.011} {\bibfield  {journal}
  {\bibinfo  {journal} {Phys. Lett. B}\ }\textbf {\bibinfo {volume} {764}},\
  \bibinfo {pages} {126} (\bibinfo {year} {2017})},\ \Eprint
  {http://arxiv.org/abs/1609.07138} {arXiv:1609.07138 [hep-ph]} \BibitemShut
  {NoStop}%
\end{thebibliography}%

\end{document}